\documentclass[pdflatex,sn-mathphys-num,referee]{sn-jnl}

\usepackage{dsfont}
\usepackage{natbib}
\usepackage{url}
\usepackage{color}
\usepackage{fullpage}
\usepackage{amsmath,amsthm,amsfonts,amssymb,amscd}
\usepackage{lastpage}
\usepackage{enumerate}
\usepackage[shortlabels]{enumitem}
\usepackage{fancyhdr}
\usepackage{mathrsfs}
\usepackage{xcolor}
\usepackage{listings}
\usepackage{dcolumn}
\usepackage{bm}
\usepackage{graphicx}
\usepackage{subcaption}
\usepackage{float}
\usepackage[normalem]{ulem}
\usepackage{hyperref}
\hypersetup{%
  colorlinks=true,
  linkcolor=blue,
  linkbordercolor={0 0 1}
}

\newcommand{\be}{\begin{equation}}
\newcommand{\ee}{\end{equation}}
\newcommand{\bea}{\begin{eqnarray}}
\newcommand{\eea}{\end{eqnarray}}
\newcommand{\nn}{\nonumber}

\newcommand{\pp}{p_{\phi}}
\newcommand{\sq}{\sqrt{q}}
\renewcommand{\H}{\mathcal{H}}
\newcommand{\lv}{\lambda_{v}}
\newcommand{\lp}{\lambda_{\phi}}
\newcommand{\sv}{\sigma_v}
\renewcommand{\sp}{\sigma_{\phi}}

\begin{document}

\title[Polymer Bianchi-I with polymer matter]{Polymer Bianchi-I with polymer matter}

\author[1]{\fnm{Aleena} \sur{Zulfiqar}}\email{aleena.706@lums.edu.pk}

\author*[2]{\fnm{Syed Moeez} \sur{Hassan}}\email{\href{mailto:syed_hassan@lums.edu.pk}{syed\_hassan@lums.edu.pk}}

\affil[1]{\orgdiv{Department of Mathematics}, \orgname{Syed Babar Ali School of Science and Engineering, Lahore University of Management Sciences}, \orgaddress{\city{Lahore}, \postcode{54792}, \country{Pakistan}}}

\affil*[2]{\orgdiv{Department of Physics}, \orgname{Syed Babar Ali School of Science and Engineering, Lahore University of Management Sciences}, \orgaddress{\city{Lahore}, \postcode{54792}, \country{Pakistan}}}

\abstract{We analyze the effective dynamics of a polymer quantized Bianchi-I universe coupled to a polymer quantized scalar field, with a pressureless dust field acting as an internal clock. We show that for a consistent polymer quantization of the anisotropies, the volume variable has to be polymerized first, and that a choice of different polymer scales for the two leads to substantially different dynamics. We further derive an effective Friedmann equation for this model, and compute the shear scalar. We find that polymer quantizing the scalar field leads to significant differences in the evolution of various cosmological quantities as compared to standard quantization. The amount of these variations, as well as the location of the quantum bounce, depends on the matter polymer scale.}

\maketitle

\section{Introduction} \label{intro}

One of the major open quests of modern physics is to gain insight into the quantum nature of spacetime. Using a background independent quantization scheme, Loop Quantum Gravity (LQG), and its cosmological manifestation Loop Quantum Cosmology (LQC), have yielded important results, including for example, a resolution of the initial big-bang singularity \cite{Agullo:2016tjh, Ashtekar:2011ni, Li:2018opr}. A method of quantization motivated from loop quantization is that of polymer quantization which introduces a fundamental discreteness in the theory \cite{Halvorson_2004, Ashtekar:2002sn}. Polymer quantization has been applied to both gravitational models \cite{BenAchour:2018jwq, Montani:2018uay, Giovannetti:2021bqh, Giovannetti:2019ewe, Barca:2019ane, Giovannetti:2020nte} (with results similar to those produced by LQC \cite{Barca:2021qdn}), and to matter systems in order to investigate the effects of such a quantization scheme on matter dynamics \cite{Ashtekar:2002vh, Corichi:2007tf, Laddha:2010hp, Kreienbuehl:2013toa, Hossain:2009ru,  Hossain:2010eb, Hassan:2014sja, Hassan:2017cje, Ali:2017fhp, Mujtaba:2025tcc}.

Although much of the existing literature has concentrated on homogeneous and isotropic spacetimes \cite{PhysRevD.81.084043, Zulfiqar:2025aef}, it is possible that the early universe (or even the late time universe) passed through anisotropic phases \cite{Schucker:2014wca, Campanelli:2006vb, Akarsu:2019pwn, Aluri:2022hzs, Hertzberg:2024uqy, Akarsu:2021max}. The Bianchi-I model, as the simplest anisotropic cosmological model, serves as a natural framework to investigate such scenarios. Studies within LQC have demonstrated that the effective dynamics of Bianchi-I spacetimes resolve the classical singularity, and that the anisotropies introduce significant modifications to the behavior of the universe in the Planckian regime \cite{Ashtekar:2009vc, deCesare:2019suk}.

From a fundamental perspective, if gravity is in fact polymer quantized, it would be natural to expect that any matter field would also follow the same quantization scheme \cite{Thiemann:1997rt}. Some models of this type have been studied before \cite{Laddha:2008em, Domagala:2012tq, Agullo_2023}, and a first detailed treatment of the effective dynamics of a homogeneous and isotropic Friedmann-Lemaitre-Robertson-Walker (FLRW) universe coupled to a polymer quantized scalar field appeared in \cite{Zulfiqar:2025aef}. In this study, we extend the model presented in \cite{Zulfiqar:2025aef} to a homogeneous but anisotropic Bianchi-I spacetime. We polymer quantize the spacetime, and couple it to a polymer quantized massless scalar field. We then study the effective dynamics arising from this model by using a pressureless dust field as an internal clock.

Effective dynamics offer a simpler way of quantifying the main quantum effects of a model without having to resort to the full quantum theory. They are obtained by computing the expectation value of the Hamiltonian in suitable `semi-classical' states \cite{Husain:2006uh, Taveras:2008ke}. And although they serve only as a starting point for a complete quantum treatment, they are surprisingly effective \cite{Rovelli:2013zaa}. These effective dynamics are then studied with respect to a dust field acting as a clock \cite{Husain:2011tk, Husain:2011tm}. A major advantage of choosing dust is that the physical (reduced) Hamiltonian is linear in the Hamiltonians of the remaining degrees of freedom (unlike for example, with scalar field time, or other clock choices \cite{Ashtekar:2011ni, Agullo:2016tjh}). The quantum dynamics arising from using dust as a clock have been investigated in detail for many different gravitational models \cite{Ali:2018vmt, Ali:2018nql, Ali:2015ftw, Ali:2017qwa, Husain:2021ojz, Giesel:2020xnb}.

The remainder of this paper is organized as follows. In Section \ref{model}, we introduce the model, outline the essentials of polymer quantization, and derive the effective Hamiltonian. The effective Friedmann equation is derived in Section \ref{eff-fried}. In Section \ref{dyn}, we examine the dynamics of this system and present numerical results for both pure gravity and gravity coupled to matter. A summary of our results and concluding remarks are provided in Section \ref{sec-con}. Throughout, we employ Planck units with $c = \hbar = 8 \pi G = 1$.

\section{Physical Hamiltonian and Bianchi-1 Model} \label{model}

We begin with General Relativity written in the Arnowitt-Deser-Misner (ADM) form, coupled to a  massless scalar field $\phi$ and a pressureless dust field $T$,
\be
\label{full-theory}
S = \int d^3x dt ~ [ \pi^{ab} \dot{q}_{ab} + p_T \dot{T} + \pp \dot{\phi}  - N \H - N^a C_a],
\ee
where,
\bea
\label{full-ham}
\H &=& \dfrac{1}{\sq} \left(\pi^{ab} \pi_{ab} - \frac{1}{2} \pi^2 \right) - \sq R \nn \\ 
   &+& \dfrac{\pp^2}{2 \sq} + \dfrac{1}{2} \sq q^{ab} \partial_a \phi \partial_b \phi \nn \\ 
   &+& p_T \sqrt{1 + q^{ab} \partial_a T \partial_b T },\\
C_a &=& D_b \pi^b_a + \pp \partial_a \phi  -p_T \partial_a T, 
\eea
are the Hamiltonian and diffeomorphism constraints respectively. $(\phi, \pp)$, $(T, p_T)$ and $(q_{ab}, \pi^{ab})$ are the scalar field, dust field and gravitational phase space variables respectively, $q$ is the determinant of the spatial metric $q_{ab}$, $\pi$ is the trace of $\pi^{ab}$, $R$ is the 3-Ricci scalar curvature, $N$ is the lapse and $N^a$ is the shift.

We reduce the full theory to a homogeneous but anisotropic Bianchi-I universe. The line element of this model is, 
\be\label{Bianchigen}
ds^2=-dt^2+a_1^2(t)dx^2+a_2^2(t)dy^2+a_3^2(t)dz^2,
\ee
where $a_1(t), a_2(t)$ and $a_3(t)$ are the scale factors corresponding to the three different coordinate directions, and the volume variable is $v=a_1a_2a_3$. From these scale factors, we can define directional Hubble parameters $H_i$ as,
\be\label{direc hubble}
H_i=\frac{\dot{a_i}}{a_i},
\ee
with the mean Hubble parameter ($H_b$) given as,
\be\label{mean hubble}
H_b = \frac{{1}}{3}\left(H_1+H_2+H_3\right).
\ee
Another variable of interest (that measures the anisotropy of spacetime) is the shear, which is defined for each direction as the corresponding Hubble parameter minus the mean Hubble parameter,\footnote{Note that we are using a slightly different notation for the shears ($\Sigma$) as compared to the literature ($\sigma$), since we use the variable $\sigma$ for state widths that appear later.}
\be \label{direc shear}
\Sigma_i=H_i-H_b
\ee
and, the shear scalar,
\be \label{shear1}
\Sigma^2=\frac{1}{3}\left(\Sigma_1^2+\Sigma_2^2+\Sigma_3^2\right).
\ee

In what follows, we work with the more convenient Misner variables defined as,
\bea \label{relation}
a_1(t)=e^\alpha e^{\beta_+ +\sqrt{3}\beta_-},~~~~~~~a_2(t)=e^\alpha e^{\beta_+ -\sqrt{3}\beta_-},~~~~~~~~
a_3(t)=e^\alpha e^{-2\beta_+},
\eea
and, the line element takes the form,
\be\label{Bianchi}
ds^2=-dt^2+e^{2\alpha}(e^{2\beta})_{ij}dx^idx^j,
\ee
where, $\alpha$ is related to the volume $v$ as $v=e^{3\alpha}$, and $\beta$ is a traceless diagonal matrix containing the two independent anisotropies $\beta_+$ and $\beta_-$ as $diag(\beta_+ +\sqrt{3}\beta_-,~ \beta_+ -\sqrt{3}\beta_-,~ -2\beta_+)$. With this ansatz, the gravitational phase space variables become,
\bea
q_{ij} &=& e^{2\alpha}\left( e^{2\beta}\right)_{ij} ~, \nn\\~~~ \pi^{ij} &=&  \frac{(p_+E_+^{ij}+p_-E_-^{ij}+q^{ij}p_{\alpha})}{2},
\eea
where $E_{+ij}=diag(1,1,-2)$, $E_{-ij}=diag(\sqrt{3},-\sqrt{3},0)$, and $(p_\alpha,~ p_\pm)$ are the conjugate momenta of $(\alpha,~\beta_\pm)$ respectively. We further impose the requirement that both the scalar and dust fields are spatially homogeneous, consistent with the homogeneity of the background universe,
\be
(\phi,~\pp ,~T ,~p_T)= (\phi(t), ~\pp(t), ~T(t), ~p_T(t)).
\ee
With this reduction, the diffeomorphism constraint vanishes identically, and the action becomes,
\be
S = V_0 \int dt ~ \Bigg( p_\alpha \dot{\alpha} + p_+\dot{\beta_+}+ p_-\dot{\beta_-}+ \pp \dot{\phi} + p_T \dot{T} - NH \Bigg),
\ee
where,
\be
\label{Ham-constr-cosm}
H = -\frac{1}{48 e^{3\alpha}}(p_\alpha^2-p_+^2-p_{-}^2) + e^{3\alpha}\Lambda+\frac{p_{\phi}^2}{2e^{3\alpha}} + p_T \approx 0
\ee
is the Hamiltonian constraint in Misner variables, and $ V_0 = \int d^3x $ is a fiducial volume.

We now choose the dust field as an internal clock, $t = -T$, and explicitly solve the classical Hamiltonian constraint to obtain a reduced physical Hamiltonian (for further details, we refer to, \cite{Husain:2011tk, Husain:2011tm}, and we are following the conventions of \cite{Hassan:2017cje}),
\be
\label{phys-ham}
H_p = p_T = \frac{1}{48 e^{3\alpha}}(p_\alpha^2-p_+^2-p_{-}^2)-e^{3\alpha}\Lambda - \frac{p_{\phi}^2}{2e^{3\alpha}}.
\ee
Since our goal will be to polymer-quantize this theory, we perform one more canonical transformation (at the classical level) to the volume variable, where the conjugate momentum to the volume is appropriately de-densitized. This is important since polymer quantization involves variables constructed by exponentiating the momenta, and therefore, the momenta must be scalars \cite{Hossain:2009ru}. The new variables are,
\bea
v =  e^{3\alpha} ~,~ p_{v}=\frac{p_\alpha}{3e^{3\alpha}}=\frac{p_\alpha}{3v},
\eea
and the physical Hamiltonian becomes,
\be
\label{Ham-bar}
H_p =H_v - H_+ - H_- - H_\phi, .
\ee
where
\bea
\label{sep-ham}
H_v&=& \frac{3}{16}vp_v^2-v\Lambda,~~~~H_+ = \frac{p_+^2}{48v},\nn\\
H_-&=&\frac{p_-^2}{48v},~~~~H_\phi=\frac{p_\phi^2}{2v}.
\eea
Our next step is to polymer quantize this Hamiltonian. We first polymer quantize the volume variable (in the next subsection), and then proceed to coupling it to the polymer quantized anisotropies and a polymer quantized scalar field. Note that there is a subtlety involved here that did not occur for the FLRW case \cite{Zulfiqar:2025aef}: the anisotropy momenta $p_\pm$, just like the scalar field momentum $\pp$, are \emph{not} scalars, and hence, directly polymer quantizing them is not reasonable. To polymer quantize the anisotropic sector, we begin by realizing that the anisotropies behave exactly (mathematically) like a massless scalar field (\ref{sep-ham}). This means that the same method that is used to polymer quantize a scalar field on a given background can be applied to the anisotropies. This is the approach we present here: the volume variable is polymer quantized, and an effective background obtained. Both the anisotropies, and the scalar field, are then polymer quantized on this effective background.\footnote{See also \cite{Aleena-prep}, where a Bianchi-I \emph{like} model is polymer quantized.}

\subsection{Polymer quantization of Bianchi-I}\label{volume}

We start with the polymer quantization of the volume variable (the $H_v$ part of the full Hamiltonian (\ref{Ham-bar})), and take the basic variables to be the volume $v$, and the exponential of the momentum conjugate to the volume $U \equiv \exp(i \lv p_v)$ which satisfy the Poisson bracket relation,
\be
\{v,U\} = i \lv U.
\ee
Polymer quantization proceeds by promoting these variables to operators on the Hilbert space $L^2(\mathbb{R}_{Bohr})$, where $\mathbb{R}_{Bohr}$ is the Bohr compactification of the real line. A basis for this Hilbert space is provided by the states $|\nu\big>$ with the inner product,
\be
\big{<}\nu|\nu'\big> = \delta_{\nu, \nu'},
\ee
where $\delta_{\nu, \nu'}$ is the generalization of the Kronecker delta to real numbers. The action of the basic operators on these states is given as,
\be
\widehat{v}  |\nu \big{>} = \nu |\nu \big{>} ~,~ \widehat{U}  |\nu \big{>}= |\nu + \lv \big{>},
\ee
with the commutator relation $[\widehat{v},\widehat{U}] = \lv \widehat{U}$. Therefore, the states $|\nu\big>$ are eigenstates of the volume operator $\widehat{v}$ with the eigenvalue $\nu$, and $\widehat{U}$ is a translation operator on these states. The operator $\widehat{U}$ is not weakly continuous in the parameter $\lv$, and hence a momentum operator does not exist. This introduces a fundamental discreteness in the theory with the scale set by $\lv$. A momentum operator can be defined indirectly however, by making use of the expansion of the classical variable $\exp(i \lv p_v)$ as follows,
\be
\widehat{p_v} \equiv \frac{1}{2i{\lambda_v}} \left( \widehat{U}-\widehat{U}^\dag \right).
\ee
From this definition of the momentum operator, the gravitational part of the Hamiltonian becomes (with a symmetric operator ordering),
\be
\widehat{H_v}=\frac{3}{16}\widehat{p_v}\widehat{v}\widehat{p_v}-\Lambda \widehat{v}.
\ee
To study the effective dynamics coming from this Hamiltonian, we compute its expectation value in a suitable `semi-classical' state (following \cite{Husain:2006uh, Taveras:2008ke}). The state is defined as,
\bea
\big{|} \psi \big{>} &=& \frac{1}{\mathfrak{N}}\sum_{-\infty}^{+\infty} A_k  \big{|}\nu_k \big{>}, \nn \\
A_k &=& \exp{\left({\dfrac{-(\nu_k-v)^2}{2\sigma_{v}^2}}\right)} \exp{({-i p_v \nu_k})},
\eea
where $(v,p_v)$ are the peaking values for this semi-classical state, $\sv$ is the gravitational state width, and $\mathfrak{N}$ is a normalization constant.\footnote{We choose a polymer lattice with $\nu_k = v + k\lv$.} The expectation value of the Hamiltonian in this state gives,
\be
\label{grav-poly-ham}
\big{<} \widehat{H}_v \big{>}=-\frac{3v}{32 \lambda_v^2} \left[ e^{-\lambda_v^2/\sigma_v^2} \cos\left( 2{\lambda_v} p_v \right)-1 \right] - \Lambda v.
\ee

With this effective background now available, we turn to the polymer quantization of the anisotropies. As stated before, for the purposes of polymer quantization, we are going to treat the anisotropies exactly like a scalar field living on this effective background. Therefore, we can use the methods first presented in \cite{Hossain:2009ru}. For completeness, we recall some relevant details here.

For a scalar field $\phi(\textbf{x})$ with momentum $P_\phi(\textbf{x})$, the standard variables of choice for polymer quantization are,
\be
\phi_f = \int d^3x \sqrt{q} f(\textbf{x}) \phi(\textbf{x}), ~~ U_\lambda = \exp \left( \dfrac{i \lambda P_\phi}{\sqrt{q}} \right),
\ee
which satisfy the Poisson bracket,
\be
\{\phi_f, U_\lambda \} = i f \lambda U_\lambda.
\ee
We note here that the factor of $\sqrt{q}$ is necessary in the definition of $U_\lambda$ to render the argument of the exponential a scalar quantity (since $P_\phi$ has a density weight 1). Similarly, this factor is necessary in the definition of $\phi_f$ to have the correct measure of integration. For the homogeneous Bianchi-I spacetime under consideration here, the smearing function $f(\textbf{x}) = 1$, and $\sqrt{q} = v$, which gives,
\be
\phi_f = v \phi, ~~ U_\lambda = \exp \left( \dfrac{i \lambda P_\phi}{v} \right).
\ee
Applying this to the anisotropy degrees of freedom, we obtain
\be
\pmb{\beta}_\pm=v\beta_\pm, ~~ U_\pm=\exp{(i\lambda_\pm p_\pm/v)}
\ee
as the basic variables (with $\lambda_\pm$ being the anisotropy polymer scales), and the anisotropy momenta operators are defined as,
\be
\widehat{p}_{\pm} \equiv \frac{v}{{2i\lambda_{\pm}}}\left(\widehat{U}_\pm-\widehat{U}_\pm^\dagger\right).
\ee

The expectation value of the anisotropy Hamiltonians ($H_\pm$) is calculated in the Gaussian state,
\bea \label{Gaussian1}
\big{|} \psi^\pm \big{>} &=& \frac{1}{\mathfrak{R'}}\sum_{-\infty}^{+\infty} A_k^\pm  \big{|}u^\pm\big{>}, \nn \\
A_k^\pm &=& \exp{\left({-\dfrac{(B^\pm-B^\pm_0)^2}{2\sigma_{\pm}^2}}\right)} \exp{(-ip_\pm B^\pm)},
\eea
where, $\mathfrak{R'}$ is a normalization constant, $(B^\pm_0,p_\pm),$ are the peaking values for the anisotropies $\beta_\pm$ and their corresponding conjugate momenta, and $\sigma_\pm$ are the state widths.\footnote{Again, we choose a polymer lattice with $B^\pm=B_0^\pm+k\lambda_\pm$.} The expectation value is computed to be,
\bea \label{H_+-}
\big{<} \widehat{H}_{\pm} \big{>} = -\frac{v}{96\lambda_\pm^2}\left( e^{ -\lambda_{\pm}^2/v^2 \sigma_\pm^2 } \cos\left(\frac{2\lambda_{\pm}p_\pm}{v}\right) -1 \right).
\eea
The total polymer quantized effective gravitational Hamiltonian becomes,
\bea\label{H_g}
H_g&=&-\frac{3v}{32 \lambda_v^2} \left( e^{-\lambda_v^2/\sigma_v^2} \cos\left( 2{\lambda_v} p_v \right)-1 \right)- \Lambda v -\frac{v}{96}\left(\frac{1}{\lambda_+^2}+\frac{1}{\lambda_-^2}\right)\nn\\ 
&+&\frac{v}{96}\left(\frac{1}{\lambda_+^2} e^{ -\lambda_{+}^2/v^2 \sigma_+^2 } \cos\left(\frac{2\lambda_{+}p_+}{v}\right)+\frac{1}{\lambda_-^2} e^{ -\lambda_{-}^2/v^2 \sigma_-^2 } \cos\left(\frac{2\lambda_{-}p_-}{v}\right)\right).
\eea

\subsection{Polymer quantized scalar field on a polymer Bianchi-I background}

We now couple a polymer quantized massless scalar field to the effective gravitational background given by the Hamiltonian (\ref{H_g}). The fundamental matter variables are chosen to be,
\be
\Phi = v\phi ~,~ U_\phi = \exp \left( \dfrac{i \lambda_{\phi} p_\phi}{v} \right),
\ee
where $\lp$ is the matter polymer scale parameter, and the variable $v$ represents the peaking value of the volume which is necessary to provide the appropriate density weight \cite{Hossain:2009ru}. The basis states are represented by $\big{|} \mu\big{>}$ with the orthogonality relation,
\be
 \big{<} \mu\big{|} \mu'\big{>} = \delta_{\mu,\mu'},
\ee
and operator action,
\be
\widehat{\Phi}  |\mu \big{>} = \mu |\mu \big{>} ~,~ \widehat{U}_\phi  |\mu \big{>}= |\mu + \lp \big{>}.
\ee
Once again, the momentum operator $\widehat{p}_\phi$ has to be defined indirectly as,
\be
\widehat{\pp} \equiv \frac{v}{2i{\lp}} \left( \widehat{U}_\phi -\widehat{U}_\phi^\dag \right).
\ee
We now compute the expectation value of the scalar field Hamiltonian ($\pp^2/2v$) in a semi-classical Gaussian state taken to be,
\bea \label{scalar-stat}
|\chi\big> &=& \frac{1}{\widetilde{\mathfrak{N}}}\sum_{-\infty}^{+\infty} C_k |\mu_k\big>, \nn \\
C_k &=& \exp \left( {\frac{-(\phi_k-\phi)^2}{ {2\sigma_{\phi}^2}}} \right)  \exp \left( -i p_\phi \phi_k \right),
\eea
where, $(\phi,\pp)$ are the peaking values for the scalar field and its conjugate momentum, $\sp$ is the state width, and $\widetilde{\mathfrak{N}}$ is a normalization factor.\footnote{Here $\phi_k=\mu_k/v$, and we choose a lattice with $\phi_k = \phi + k \lp$.} The expectation value is computed to be,
\be
\label{hphi-eff}
\big{<} \widehat{H}_{\phi} \big{>} = -\frac{v}{4\lambda_\phi^2}\left( e^{ -\lambda_{\phi}^2/v^2 \sigma_\phi^2 } \cos\left(\frac{2\lambda_{\phi}p_\phi}{v}\right) -1 \right).
\ee
The total physical Hamiltonian (\ref{Ham-bar}) is the sum of the gravitational (\ref{H_g}) and the matter (\ref{hphi-eff}) Hamiltonians,
\bea
\label{poly-phys-ham}
H_p &=& -\frac{3v}{32 \lambda_v^2}  e^{-\lambda_v^2/\sigma_v^2} \cos\left( 2{\lambda_v} p_v \right) - v \left( \Lambda -\frac{3}{32 \lambda_v^2} +\frac{1}{96 \lambda_+^2}+\frac{1}{96 \lambda_-^2} +\frac{1}{4\lambda_\phi^2} \right) \nn\\ 
&+&\frac{v}{96}\left[\frac{1}{\lambda_+^2} e^{ -\lambda_{+}^2/v^2 \sigma_+^2 } \cos\left(\frac{2\lambda_{+}p_+}{v}\right)+\frac{1}{\lambda_-^2} e^{ -\lambda_{-}^2/v^2 \sigma_-^2 } \cos\left(\frac{2\lambda_{-}p_-}{v}\right)\right]\nn\\
&+&\frac{v}{4\lambda_\phi^2} e^{ -\lambda_{\phi}^2/v^2 \sigma_\phi^2 } \cos\left(\frac{2\lambda_{\phi}p_\phi}{v}\right).
\eea

\section{Effective Friedmann Equation} \label{eff-fried}

Before proceeding to discuss the effective dynamics generated by the physical Hamiltonian above, we reformulate this model into an effective Friedmann equation. Note that our main emphasis in the rest of the article will \emph{not} be on this equation for two primary reasons: (i) We are working in a fixed time gauge, and have (strongly) solved the Hamiltonian constraint at the classical level. Therefore the constraint does not exist anymore; and (ii) When writing the Friedmann equation, matter contributions (and in particular various matter parameters) appear obscured by referring only to the energy densities, whereas we want to explicitly refer to, and discuss the effects of, the matter polymer scale $\lp$ on the dynamics.

Nevertheless, since most of the literature does discuss the effective Friedmann equation, we provide this description here. To begin with, we first derive the Friedmann equation for the classical Bianchi-I model in the Misner variables we consider here. This will help clarify the process, and set the stage for deriving it in the polymer case.

To begin, we have to rewrite the Hamiltonian constraint ($\mathcal{H} = - H_P + P_T \approx 0$) as the Friedmann equation (in terms of the mean Hubble parameter $H_b$, and the matter energy densities). The mean Hubble parameter is $H_b=\dot{v}/3v$, the energy density corresponding to the cosmological constant is $\rho_{\Lambda} = \Lambda$, the energy density of the dust field is $\rho_D = P_T/v$ (since $P_T$ is the dust Hamiltonian density), and the scalar field energy density is $\rho_{\phi} = H_{\phi}/v$ where $ H_{\phi} $ is given in (\ref{sep-ham}). Since the anisotropies $\beta_\pm$ act as scalar fields, we can define an equivalent energy density for them as, $\rho_{\pm} \equiv H_{\pm}/v$ (with $ H_{\pm} $ in (\ref{sep-ham})). This gives the total `matter' energy density (including contributions from the anisotropies and the cosmological constant) as, $\rho_\text{tot} =\rho_+ +\rho_- + \rho_{\phi} + \rho_D + \rho_{\Lambda}$.

Solving the Hamiltonian constraint for the mean Hubble parameter (after substituting $\dot{v}$ in terms of $p_v$ from its equation of motion) then gives,
\be
H_b^2=\frac{1}{12}\rho_\text{tot}.
\ee
We can also compute how the shear scalar is related to the anisotropy energy densities, and find that,
\be
\label{class-shear}
\Sigma^2=\frac{1}{6}(\rho_+ +\rho_-).
\ee
In terms of this shear scalar, the Friedmann equation can be recast into a more standard form,
\be
H_b^2=\frac{1}{12}\rho_\text{tot} = \frac{1}{12} \left( \rho_+ +\rho_- + \rho_{\phi} + \rho_D + \rho_{\Lambda} \right) = \frac{1}{12} \left( \rho_+ +\rho_- + \rho_M \right) = \frac{1}{12} \rho_M + \frac{1}{2} \Sigma^2,
\ee
where we have defined $\rho_M$ as the sum of all matter contributions.

We now repeat the same steps for the polymer case, with the difference being that, the scalar field energy density is given by $\big{<} \widehat{\rho}_{\phi} \big{>} = \big{<} \widehat{H}_{\phi} \big{>}/v$ (with $\big{<} \widehat{H}_{\phi} \big{>}$ given in (\ref{hphi-eff})), and the effective anisotropy energy densities are given as $\big{<} \widehat{\rho}_{\pm} \big{>} = \big{<} \widehat{H}_{\pm} \big{>}/v$ (with $\big{<} \widehat{H}_{\pm} \big{>}$ given in (\ref{H_+-})). This gives the total `matter' energy density as, $\rho_\text{tot} = \big{<} \widehat{\rho}_{+} \big{>}+\big{<} \widehat{\rho}_{-} \big{>} + \big{<} \widehat{\rho}_{\phi} \big{>} + \rho_D + \rho_{\Lambda}$. Solving the Hamiltonian constraint then gives us (with $\tilde{\rho} = 3/(16 \lv^2)$),
\be
H_b^2 = \frac{\rho_\text{tot}}{12} \left( 1 - \frac{\rho_\text{tot}}{\tilde{\rho}} \right) + \dfrac{e^{-2 \lv^2/ \sv^2} - 1}{16^2 \lv^2}.
\ee

This can be cast into the standard effective Friedmann equation form if we define a fictitious energy density associated with the semiclassical state width as,\footnote{Note that $\rho_{\sigma} \rightarrow 0$ as the state becomes sharply peaked ($\sv \rightarrow \infty$).}
\be
\rho_{\sigma} \equiv \dfrac{3 (e^{ -\lv^2/ \sv^2} - 1)}{32 \lv^2}.
\ee
The total energy density is now given as $\tilde{\rho}_\text{tot} = \rho_\text{tot} + \rho_\sigma$, and we get,
\be
H_b^2 = \dfrac{e^{ -\lv^2/ \sv^2}}{12} \tilde{\rho}_\text{tot} \left(1 - \dfrac{\tilde{\rho}_\text{tot}}{\rho_c} \right),
\ee
with,
\be
\rho_c = \dfrac{3}{16 \lambda^2} e^{ -\lv^2/ \sv^2}
\ee
being the critical energy density.

We note that while there is no effective Friedmann equation for loop quantized Bianchi-I models \cite{McNamara:2022dmf, Motaharfar:2023hil}, one can be derived for the polymer quantization scheme presented here, and our results are comparable to those found in \cite{Giovannetti:2021bqh} (we recently became aware of \cite{Barca:2025fpu}, where the authors carry out a quantization of Bianchi-I (and Bianchi-II) with a deformed commutation relation, and also report a similar effective Friedmann equation. See also \cite{Barca:2021epy} for a comparison of the two approaches).

We can also compute the relation between the shear scalar and the effective anisotropy energy densities,
\bea
\label{poly-shear}
\Sigma^2 &=& \frac{\big{<} \widehat{\rho}_{+} \big{>}}{6} \left( 1 - \frac{\big{<} \widehat{\rho}_{+} \big{>}}{\tilde{\rho}_+} \right) + \frac{\big{<} \widehat{\rho}_{-} \big{>}}{6} \left( 1 - \frac{\big{<} \widehat{\rho}_{-} \big{>}}{\tilde{\rho}_-} \right) \nn \\
&+& f_+(\lambda_+, \sigma_+, v) + f_-(\lambda_-, \sigma_-, v),
\eea
with,\footnote{$f_\pm \rightarrow 0$ as $\sigma_\pm \rightarrow \infty$ (\emph{independently} of the polymer scales $\lambda_\pm$). In the limit that the polymer scales vanish $\lambda_\pm \rightarrow 0$ but the state widths $\sigma_\pm$ remain finite, $f_\pm \sim \mathcal{O}(1/v^2 \sigma_\pm^2)$.}
\be
f_{\pm} \equiv \dfrac{e^{-2 \lambda_{\pm}^2/v^2 \sigma_{\pm}^2} - 1}{48^2 \lambda_{\pm}^2},
\ee
and,
\be
\tilde{\rho}_{\pm} = \frac{1}{48 \lambda_{\pm}^2}.
\ee

\section{Effective Dynamics}\label{dyn} 

We now turn to describing the dynamics generated (in dust time) by the effective physical Hamiltonian (\ref{poly-phys-ham}), which contains the effects of polymer quantization in both the gravitational and the matter sectors. The equations of motion are (where an overdot represents differentiation with respect to dust time),
\bea \label{gra+sca}
\dot{v}&=&\frac{3v}{16\lambda_v}e^{-\lambda_v^2/\sigma_v^2}\sin{(2\lambda_v p_v)},\nn \\
\dot{\beta}_{\pm}&=&-\frac{1}{48\lambda_\pm}e^{-\lambda_\pm^2/v^2 \sigma_\pm^2}\sin{\left(\frac{2\lambda_\pm p_\pm}{v}\right)}, ~~ \dot{p}_{\pm}=0, \nn \\
\dot{\phi}~~&=&-\frac{1}{2\lambda_{\phi}} e^{ -\lambda_{\phi}^2/v^2 \sigma_\phi^2 }\sin\left(\frac{2\lambda_{\phi}p_\phi}{v}\right), ~~ \dot{p}_{\phi} = 0, \nn \\
\dot{p}_{v}&=&\frac{3}{32\lambda_v^2}e^{-\lambda_v^2/\sigma_v^2}\cos{(2\lambda_v p_v)} +\frac{1}{96}\left(\frac{1}{\lambda_+^2}+\frac{1}{\lambda_-^2}\right)-\frac{3}{32\lambda_v^2}+\frac{1}{4\lambda_\phi^2}+\Lambda\nn \\ 
&-&e^{-\lambda_+^2/v^2\sigma_+^2}\left[\cos{\left(\frac{2\lambda_+ p_+}{v}\right)}\left(\frac{1}{48v^2\sigma_+^2}+\frac{1}{96\lambda_+^2}\right)+\sin{\left(\frac{2\lambda_+ p_+}{v}\right)}\left(\frac{p_+}{48v\lambda_+}\right)\right]\nn\\
&-&e^{-\lambda_-^2/v^2\sigma_-^2}\left[\cos{\left(\frac{2\lambda_- p_-}{v}\right)}\left(\frac{1}{48v^2\sigma_-^2}+\frac{1}{96\lambda_-^2}\right)+\sin{\left(\frac{2\lambda_- p_-}{v}\right)}\left(\frac{p_-}{48v\lambda_-}\right)\right]\nn \\
&-&e^{ -\lambda_{\phi}^2/v^2 \sigma_\phi^2 }\left[\cos\left(\frac{2\lambda_{\phi}p_\phi}{v}\right) \left(  \frac{1}{2v^2\sigma_{\phi}^2}+\frac{1}{4\lambda_\phi^2} \right)
+ \sin\left(\frac{2\lambda_{\phi}p_\phi}{v}\right) \left( \frac{p_{\phi}}{2v\lambda_{\phi}} \right)  \right],
\eea
and, the mean Hubble parameter $H_b$ is given as,
\bea
H_b=\frac{\dot{v}}{3v}=\frac{1}{16\lambda_v}e^{-\lambda_v^2/\sigma_v^2}\sin{(2\lambda_v p_v)}.
\eea
We perform numerical evolutions for various sets of initial conditions $(v, p_v,\beta_+,p_+,\beta_-,p_-, \phi, \pp)|_{t=0}$, and for different parameter choices $(\lv, \lambda_+, \lambda_-, \lambda_\phi, \sigma_v, \sigma_+, \sigma_-, \sigma_\phi)$. The initial data are selected so that the physical Hamiltonian (\ref{poly-phys-ham}) is strictly non-negative. Moreover, because the physical Hamiltonian in dust time (\ref{poly-phys-ham}) is a constant of motion (time-independent), the dust energy density remains non-negative throughout the entire evolution. Let us first note some salient features of the above equations of motion:\\

(i) In the Bianchi-I model, since there is no potential term, the momenta conjugate to the anisotropies $(p_+,p_-)$ are constants of motion. Similarly, since we are considering only a massless scalar field, the scalar field momentum $p_\phi$ is also a constant of motion.\\

(ii) The parameters $\sigma_v, \sigma_+, \sigma_-, \sigma_\phi$ represent corrections due to finite width of the gravitational and matter Gaussian states. These corrections go to zero in the limit that the states are sharply peaked \cite{Husain:2006uh, Taveras:2008ke}. \\

(iii) Polymer quantization effects are determined by the parameters $(\lv, \lambda_+, \lambda_-, \lambda_\phi)$. The limit $\lp \rightarrow 0$ corresponds to a standard scalar field living on a polymer quantized Bianchi-I background, and the simultaneous limits $\lv, \lambda_\pm \rightarrow 0$ reproduce a polymer quantized scalar field living on a classical Bianchi-I background.\footnote{Given that we are taking three different polymer scales for the gravitational variables, we could also consider `strange' limits where one of them vanishes but the others do not. For example, we could take $\lv \rightarrow 0$ while simultaneously keeping $\lambda_\pm$ fixed at some non-zero value. This would represent a scenario where the volume variable behaves classically, while the anisotropies exhibit polymer quantization effects. While an interesting mathematical curiosity, such limits do not seem physically reasonable.} \\

(iv) The last terms in the first line of the Hamiltonian (\ref{poly-phys-ham}) show that all of the different polymer scales contribute to the total cosmological constant term. The two anisotropy polymer scales and the scalar field polymer scale contribute positively to $\Lambda$, while the volume polymer scale gives a negative contribution. This means that there can be a situation where the overall effective cosmological constant becomes negative, and therefore produces an oscillating universe with multiple bounces and re-collapses. Since we want to focus on the purely quantum effects, we have included a standard $\Lambda$ term, and given the values of the other polymer scales for various simulations, we choose a value for $\Lambda$ that does not lead to the overall term being negative.

\subsection{Effective Dynamics Of Bianchi-1 in dust time}

We first turn to describing the effective polymer dynamics of the gravitational sector, and then in the later subsection, will look at the effective dynamics with the scalar field included.

\begin{figure}[h]
\begin{center}
\includegraphics[scale=0.4]{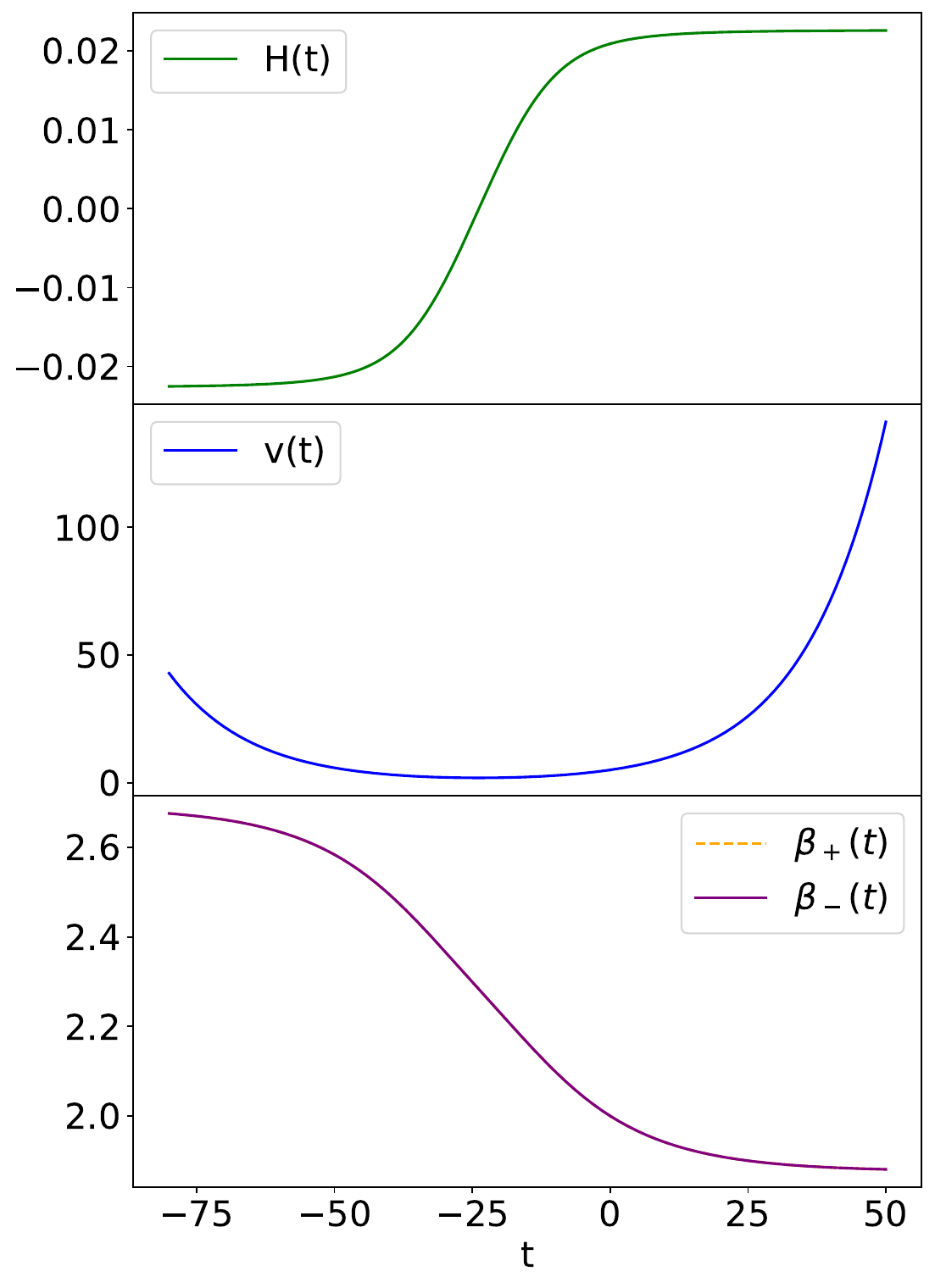}
\end{center}
\caption{\label{fig-gvaryh02} Results of a typical simulation showing the (dust) time evolution of the Hubble parameter, volume and anisotropies. Both anisotropies show same dynamics since the initial conditions are kept the same.}
\end{figure}

\begin{figure}[H]
    \centering
    \begin{subfigure}{0.4\textwidth}
        \centering
        \includegraphics[width=\linewidth]{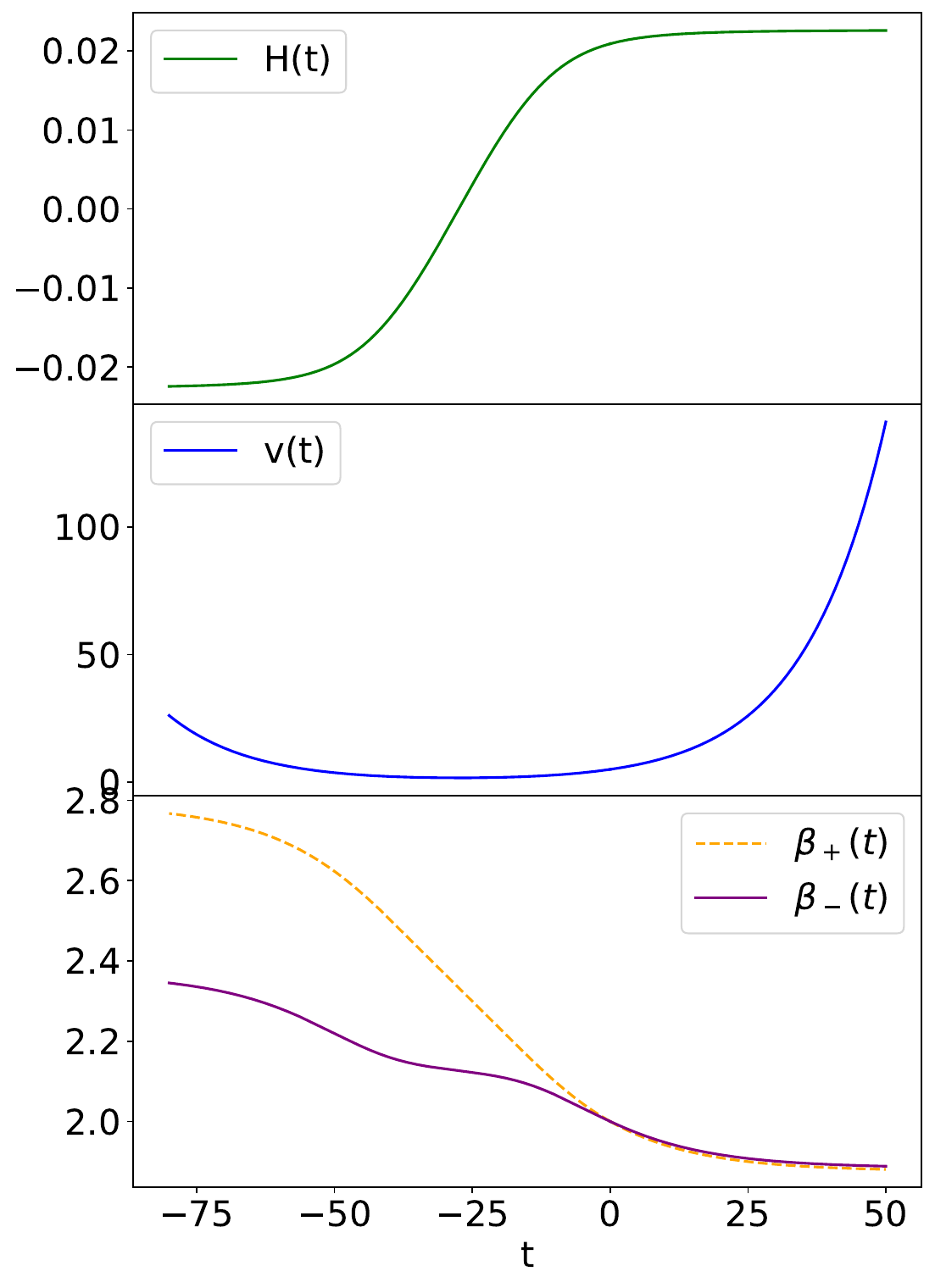}
        \caption{The two anisotropy polymer scales are varied with $\lambda_+=1$, and $\lambda_-=2$.}
        \label{fig:hubble}
    \end{subfigure}
    \begin{subfigure}{0.4\textwidth}
        \centering
        \includegraphics[width=\linewidth]{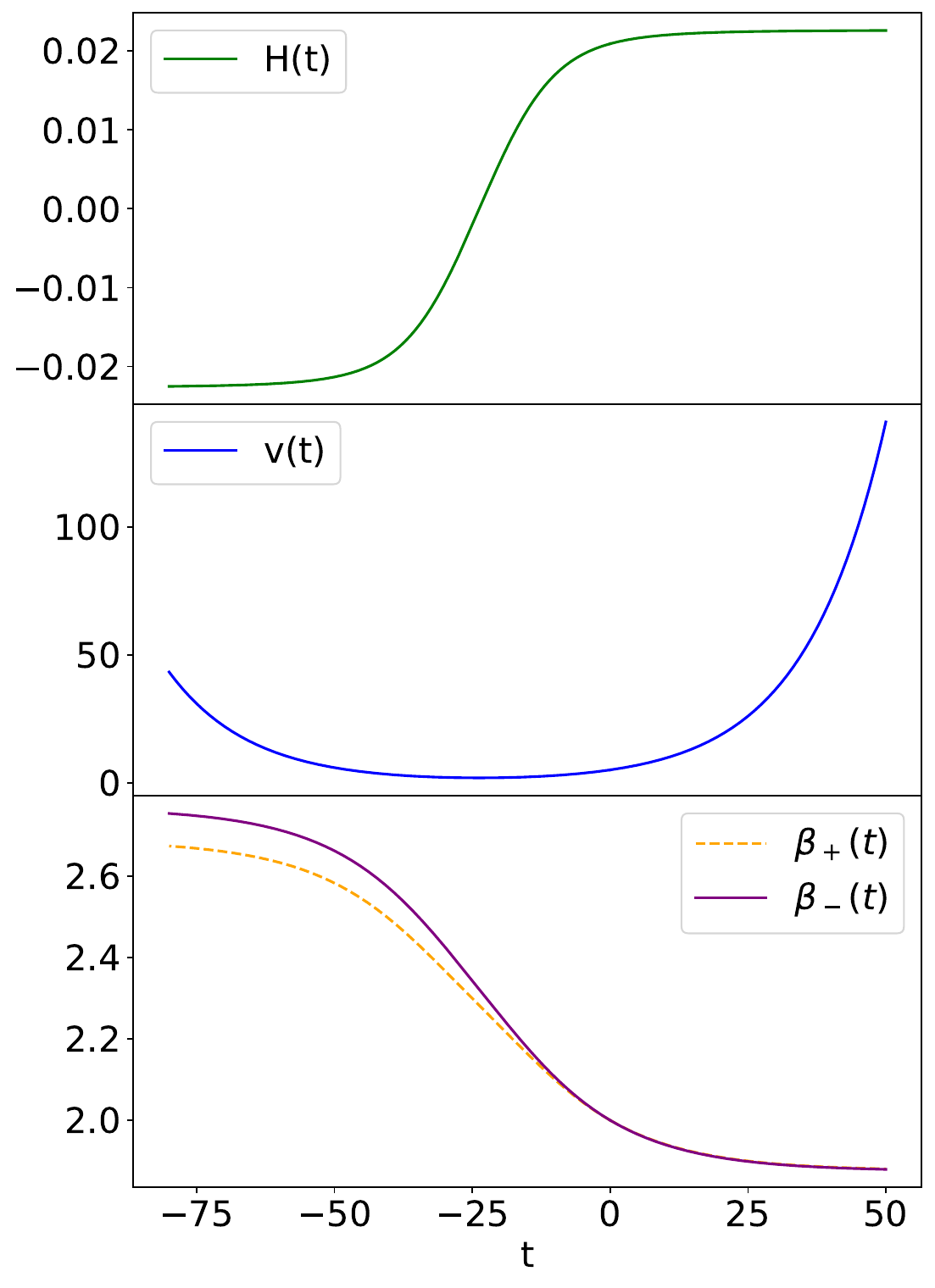}
        \caption{The two anisotropy state widths are varied with $\sigma_+=1$, and $\sigma_-=2$.}
        \label{fig:volume}
    \end{subfigure}

    \par\medskip

    \begin{subfigure}{0.4\textwidth}
        \centering
        \includegraphics[width=\linewidth]{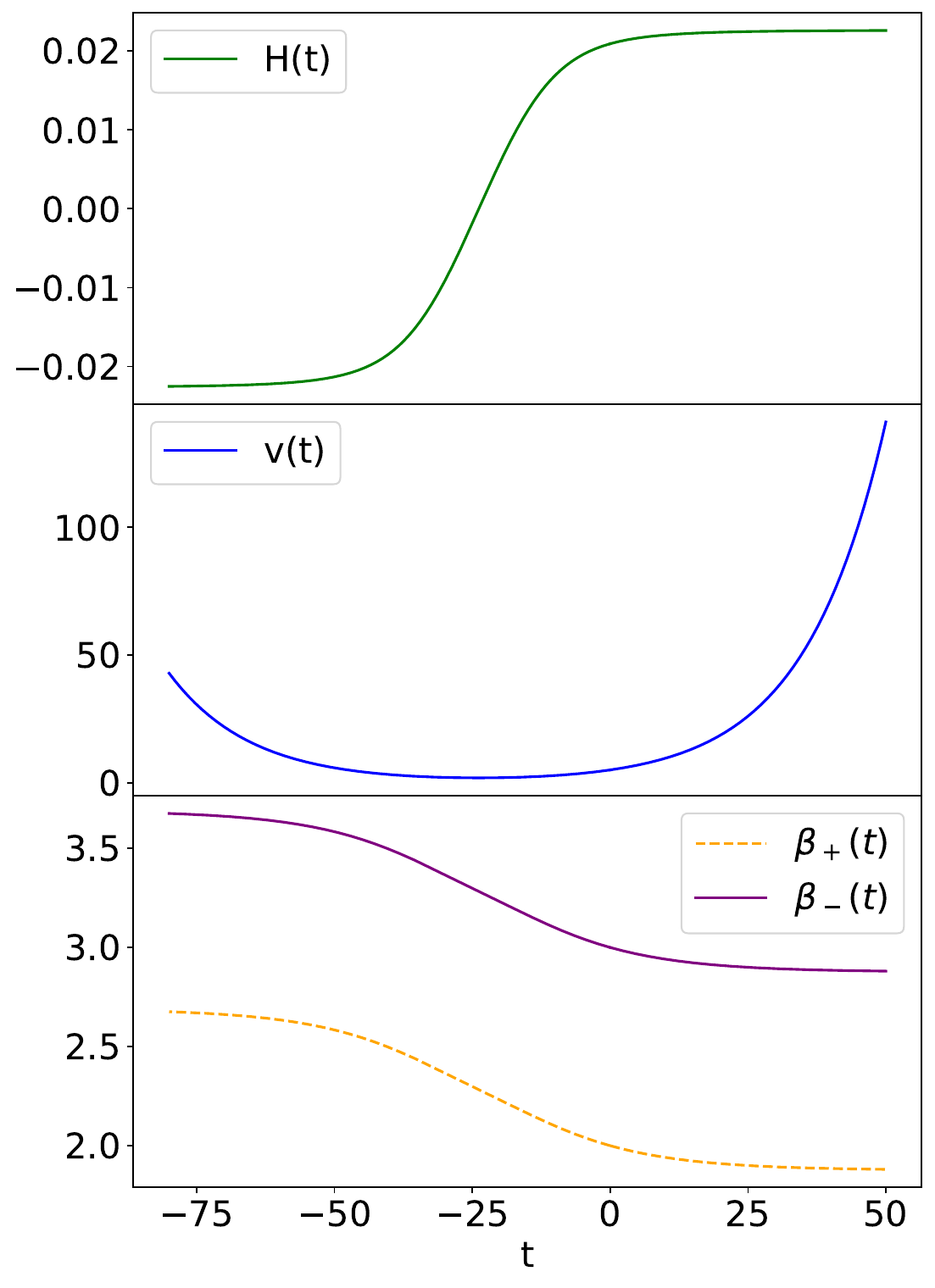}
        \caption{The anisotropy initial conditions are varied with $(\beta_+,\beta_-)|_{t=0}=(2,3)$.}
        \label{fig-gvaryh0}
    \end{subfigure}
    \begin{subfigure}{0.4\textwidth}
        \centering
        \includegraphics[width=\linewidth]{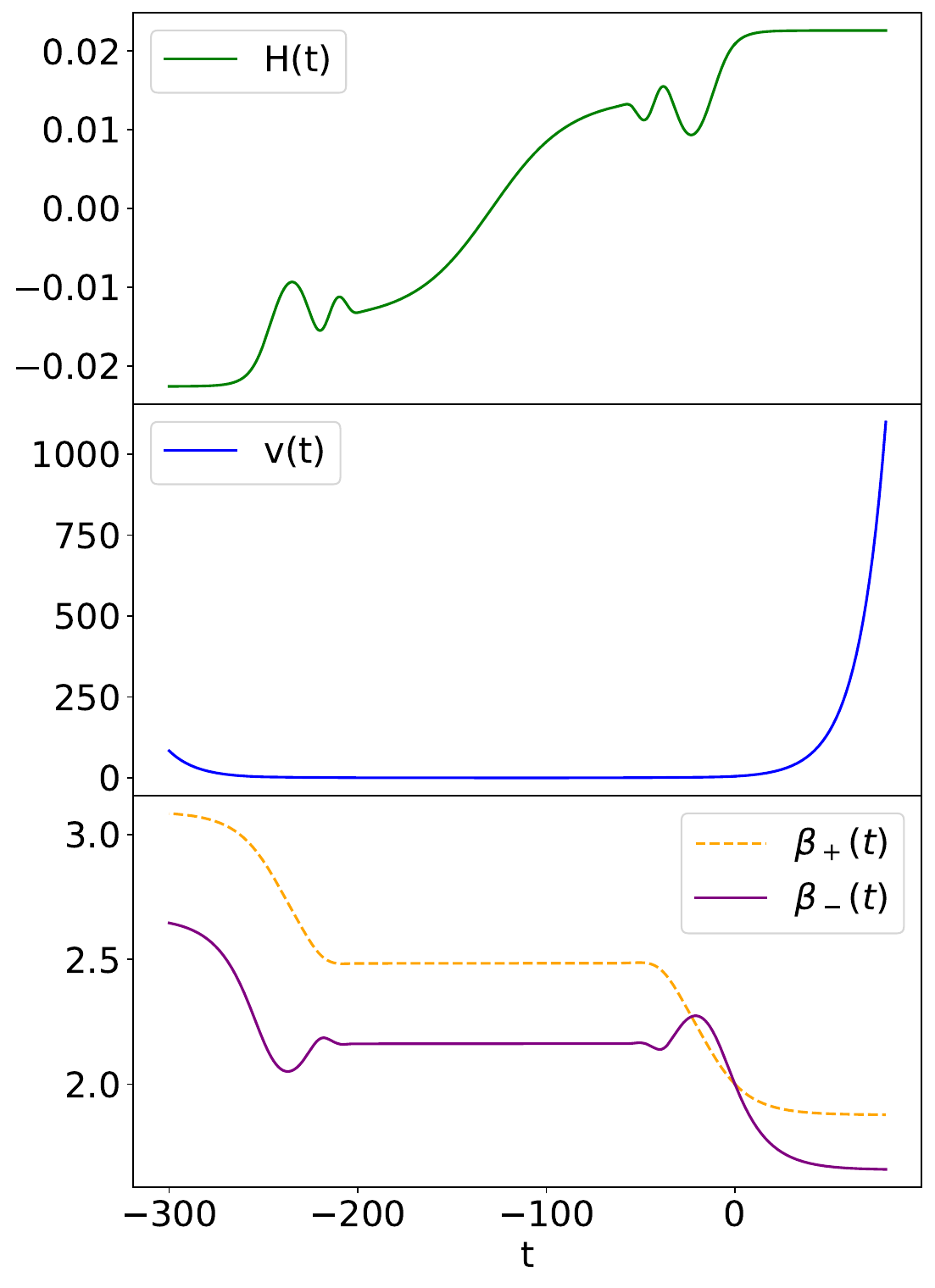}
        \caption{The two anisotropy momenta are varied with $p_+ = 1$, and $p_- = 3.$}
        \label{fig-gvaryh03}
    \end{subfigure}

    \caption{Evolution of Bianchi-I cosmological quantities with varying parameter values and initial conditions.}
    \label{fig:all_three}
\end{figure}

Figure \ref{fig-gvaryh02} shows the results of a typical simulation, and Figure \ref{fig:all_three} shows the effects of changing various parameters and initial conditions. The initial conditions for both of these figures were chosen to be $(v,P_v)|_{t=0} = (5, 1)$, and $(\beta_\pm, p_\pm) = ( 2, 1)$, and the parameter values were taken to be $(\lambda_v, \lambda_+, \lambda_-) = (1,1,1),$ $\Lambda=0.1$, and $(\sigma_v, \sigma_+, \sigma_-) = (1,1,1)$ (in Figure \ref{fig:all_three}, where various parameters and initial conditions are varied, the captions indicate which variable was changed).

Figure \ref{fig-gvaryh02} shows that when all anisotropy parameters are set equal with identical initial conditions for the corresponding variables, both anisotropies have the same evolution. The dynamics of the volume variable indicates a bouncing universe with expansion taking place before and after the bounce (as observed in \cite{Zulfiqar:2025aef} for an FLRW universe).

In Figure \ref{fig:all_three}, we illustrate how the evolution of anisotropies with respect to dust time changes when varying the polymer scales, state widths, and initial data. This change in parameters also affects the evolution of the volume variable, and in particular, there is a change in the location of the quantum bounce. We also note an interesting feature in the last panel of this figure that the Hubble parameter, and one of the anisotropies undergo oscillations before and after the bounce. These oscillations are a direct consequence of polymer quantization of the anisotropies. Specifically, the equation of motion for the anisotropies (\ref{gra+sca}) behaves as $\dot{\beta}_{\pm} \sim e^{-1/v^2} \sin(2p_{\pm}/v)$. It is this trigonometric factor (a characteristic feature of polymer quantization) that produces turning points in the evolution of the anisotropies, and hence causes an oscillatory behavior. Since $v$ is bounded below, and there is an additional exponential factor in the equation of motion, these oscillations only become significant for large numerical values of the anisotropy momenta $p_{\pm}$ (which is why they are not apparent in the other panels of this figure, or for the second anisotropy in the last panel). A similar analysis holds for the Hubble parameter where oscillations occur due to the dynamics of $p_v$ -- a quantity whose equation of motion also contains the exponential and trigonometric factors discussed above.

\subsection{Polymer Bianchi-I with a polymer scalar field}

\begin{figure}[H]
 \centering
 \begin{subfigure}{0.33\textwidth}
   \centering
   \includegraphics[width=\linewidth]{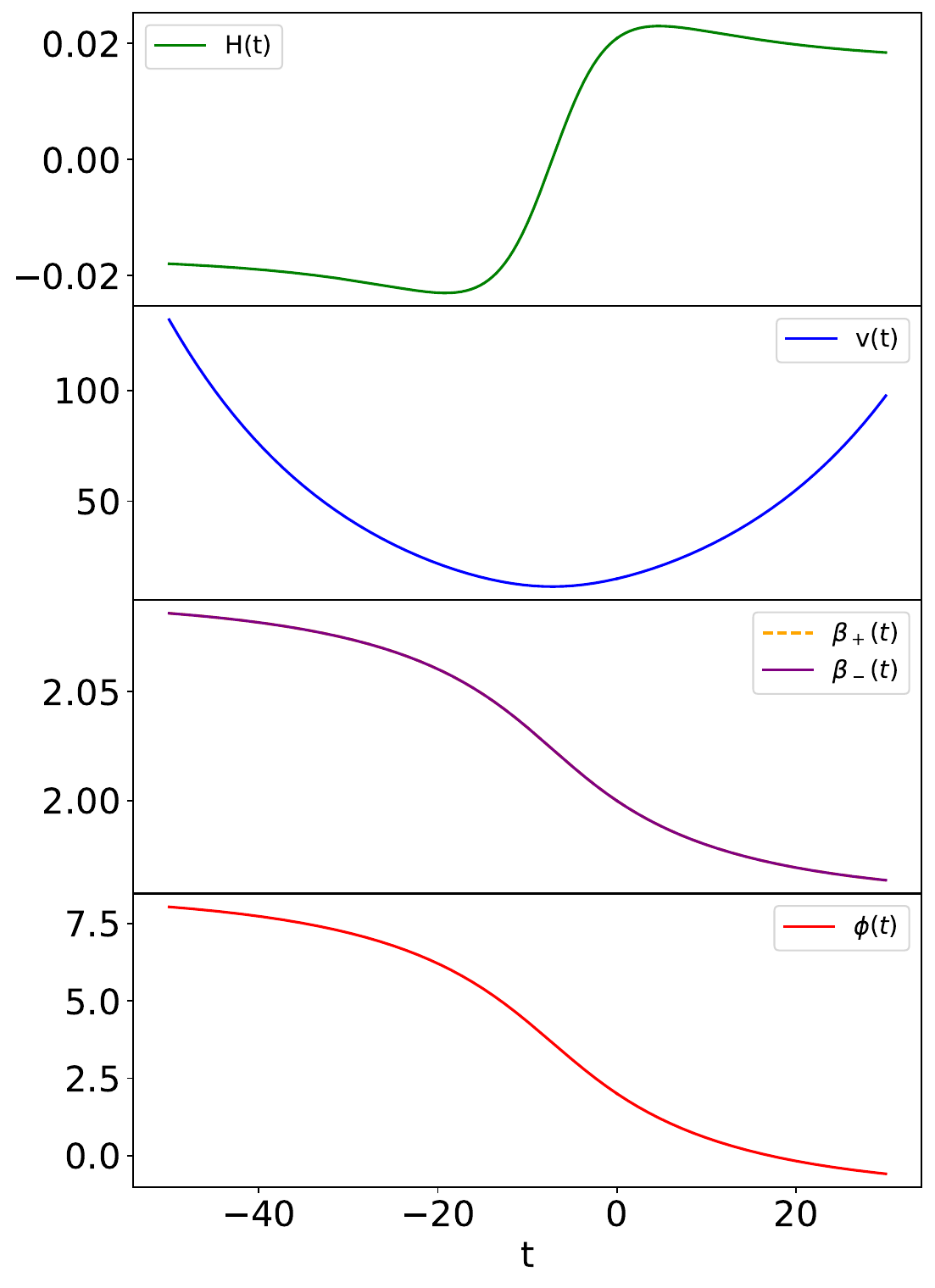}
   \caption{\label{fig-gvaryhp02} Same parameter values and initial conditions.}
 \end{subfigure}%
 \hfill
 \begin{subfigure}{0.33\textwidth}
   \centering
   \includegraphics[width=\linewidth]{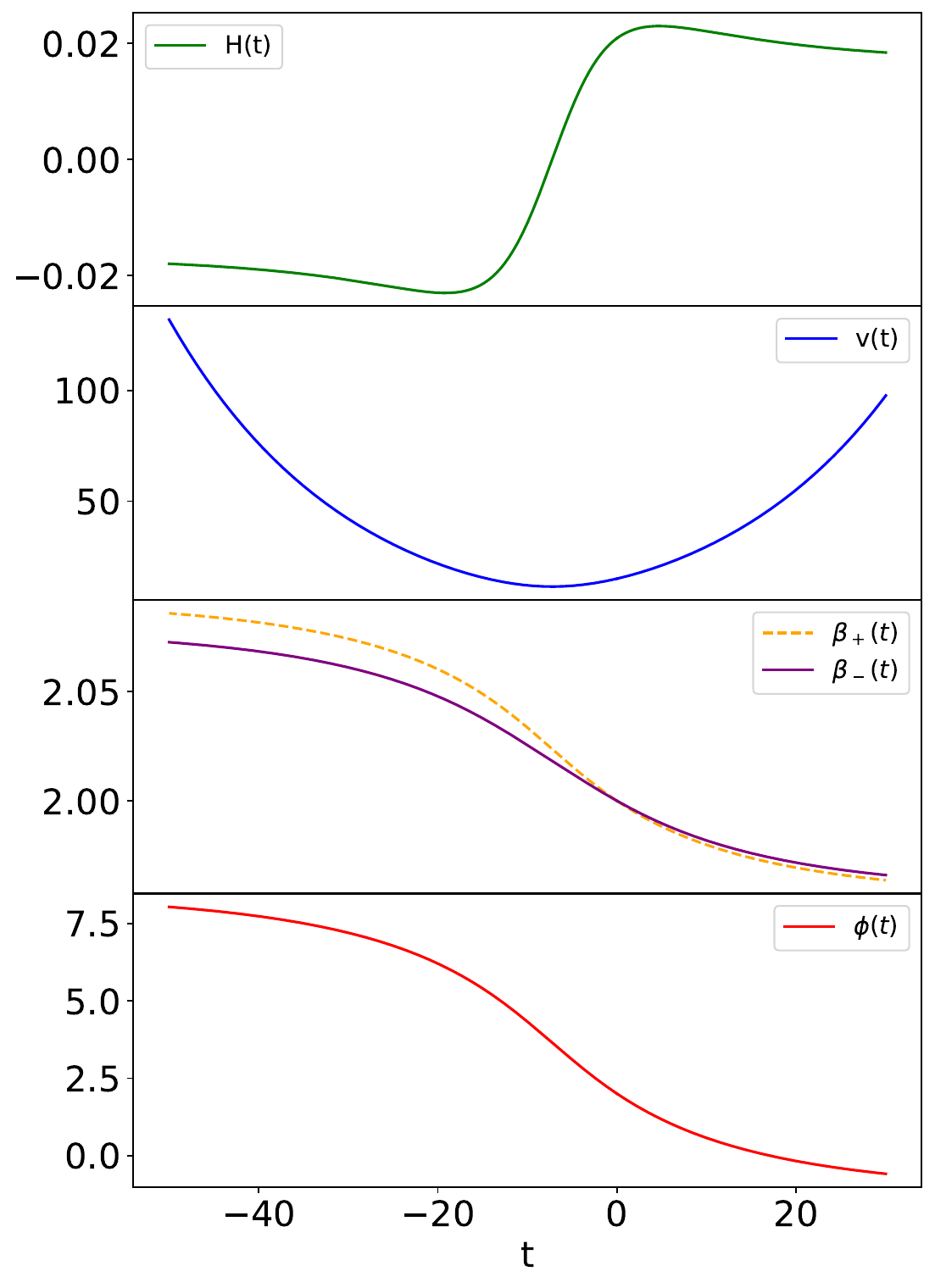}
   \caption{Anisotropy polymer scales are varied: $\lambda_+=1$, $\lambda_-=5$.}
   \label{fig:hubblep}
 \end{subfigure}%
 \hfill
 \begin{subfigure}{0.33\textwidth}
   \centering
   \includegraphics[width=\linewidth]{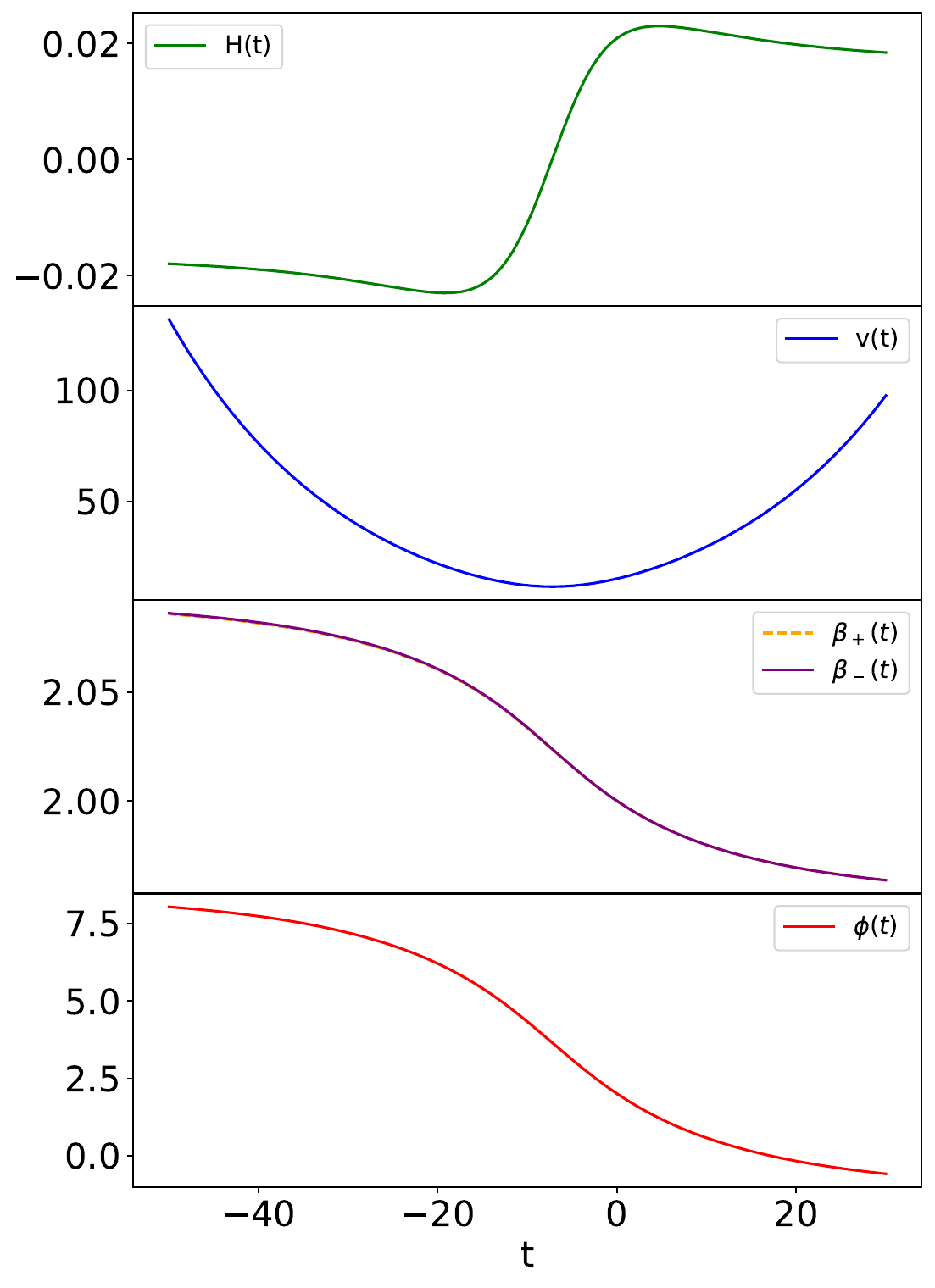}
   \caption{The anisotropy state widths are varied with $\sigma_+=1$, and $\sigma_-=2$.}
   \label{fig:volumep}
 \end{subfigure}

 \par\medskip

 \begin{subfigure}{0.33\textwidth}
   \centering
   \includegraphics[width=\linewidth]{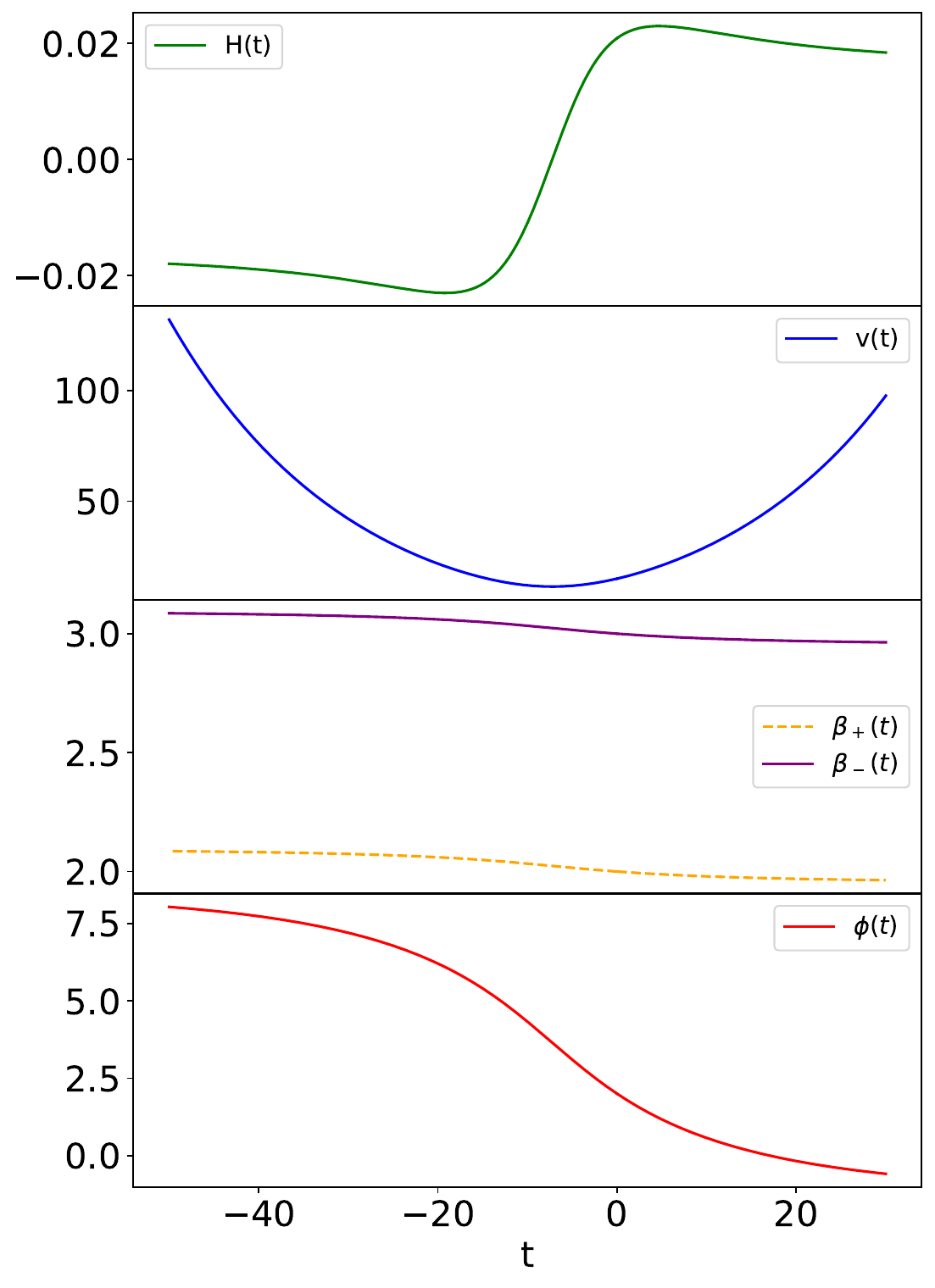}
   \caption{Anisotropy initial conditions are varied: $(\beta_+,\beta_-)|_{t=0}=(2,3)$.}
   \label{fig-gvaryhp0}
 \end{subfigure}%
 \hfill
 \begin{subfigure}{0.33\textwidth}
   \centering
   \includegraphics[width=\linewidth]{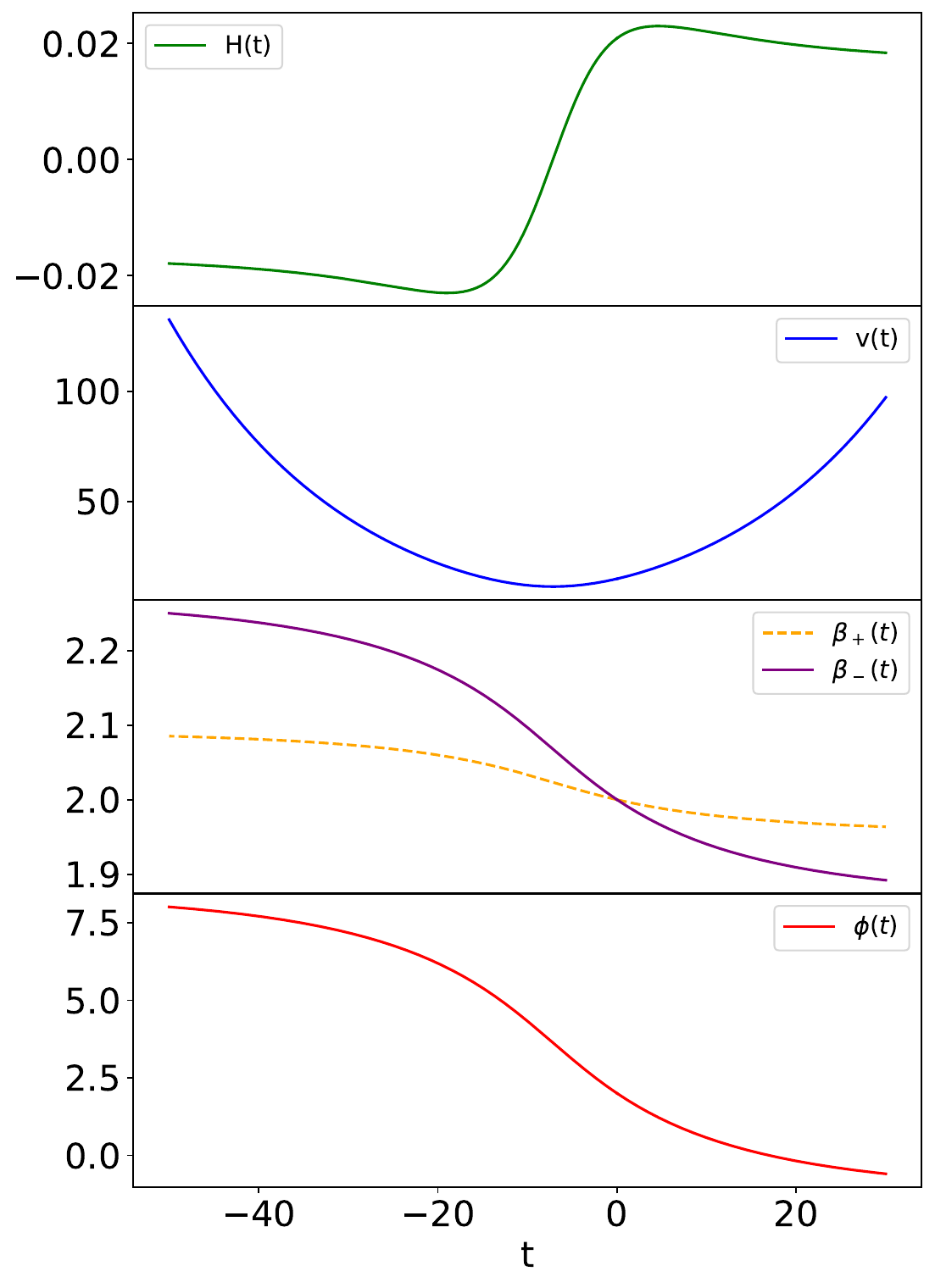}
   \caption{The anisotropy momenta are varied with $(p_+,p_-)=(1,3).$}
   \label{fig-gvaryhp01}
 \end{subfigure}%
 \hfill
 \begin{subfigure}{0.33\textwidth}
   \centering
   \includegraphics[width=\linewidth]{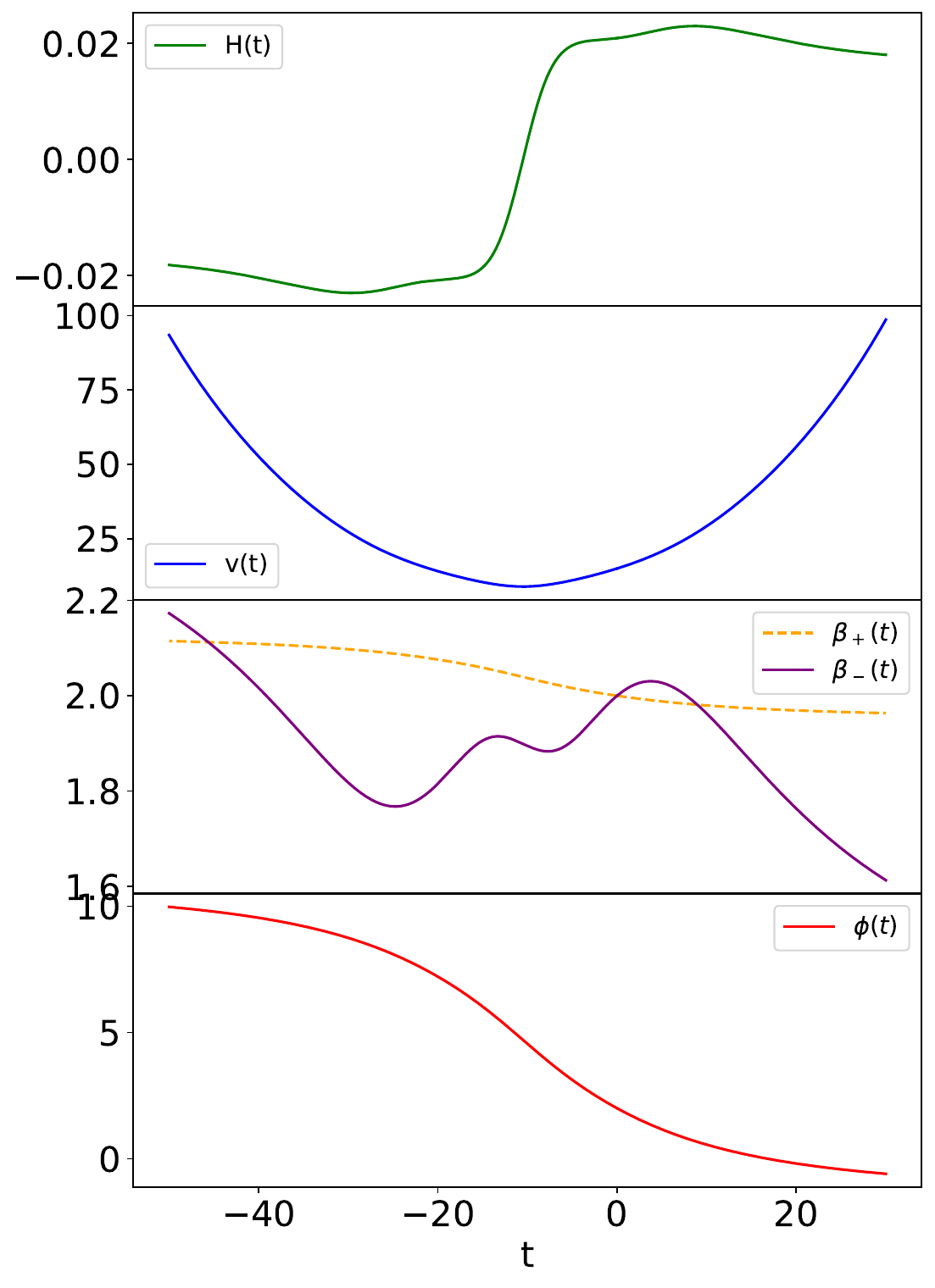}
   \caption{The anisotropy momenta are varied with $(p_+,p_-)=(1,30).$}
 \end{subfigure}

 \caption{Time evolution of cosmological quantities in Bianchi-I with varying parameter values and initial conditions.}
 \label{fig:all_three1}
\end{figure}

\begin{figure}[h]
    \centering
        \includegraphics[scale=0.3]{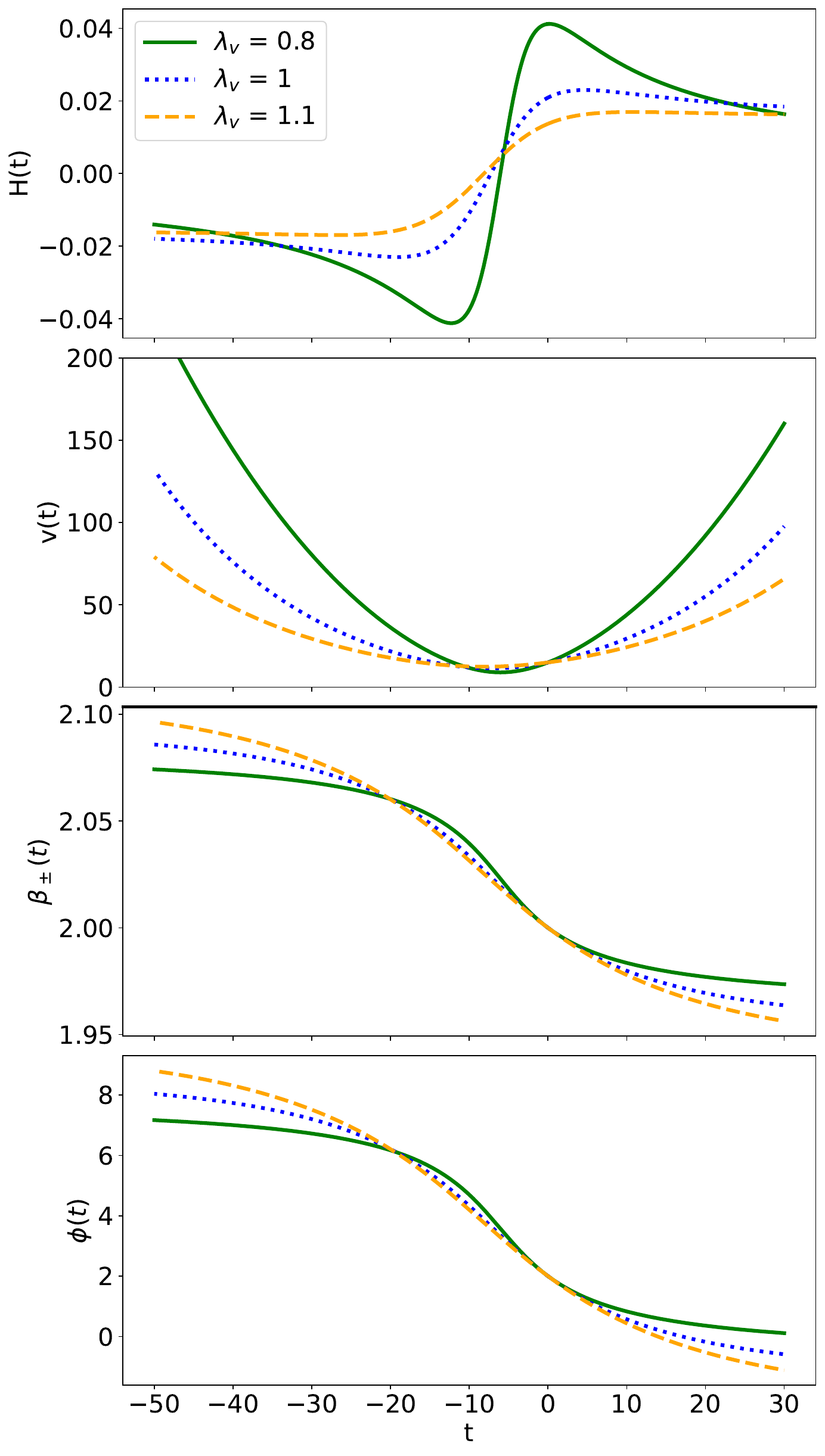}
 \caption{The effects of varying the volume polymer scale $\lambda_v$ on the evolution of various cosmological quantities and the scalar field. The differences are clearly visible.}
 \label{fig-all-lambda-v}
\end{figure}

\begin{figure}[h]
 \centering
        \includegraphics[scale=0.3]{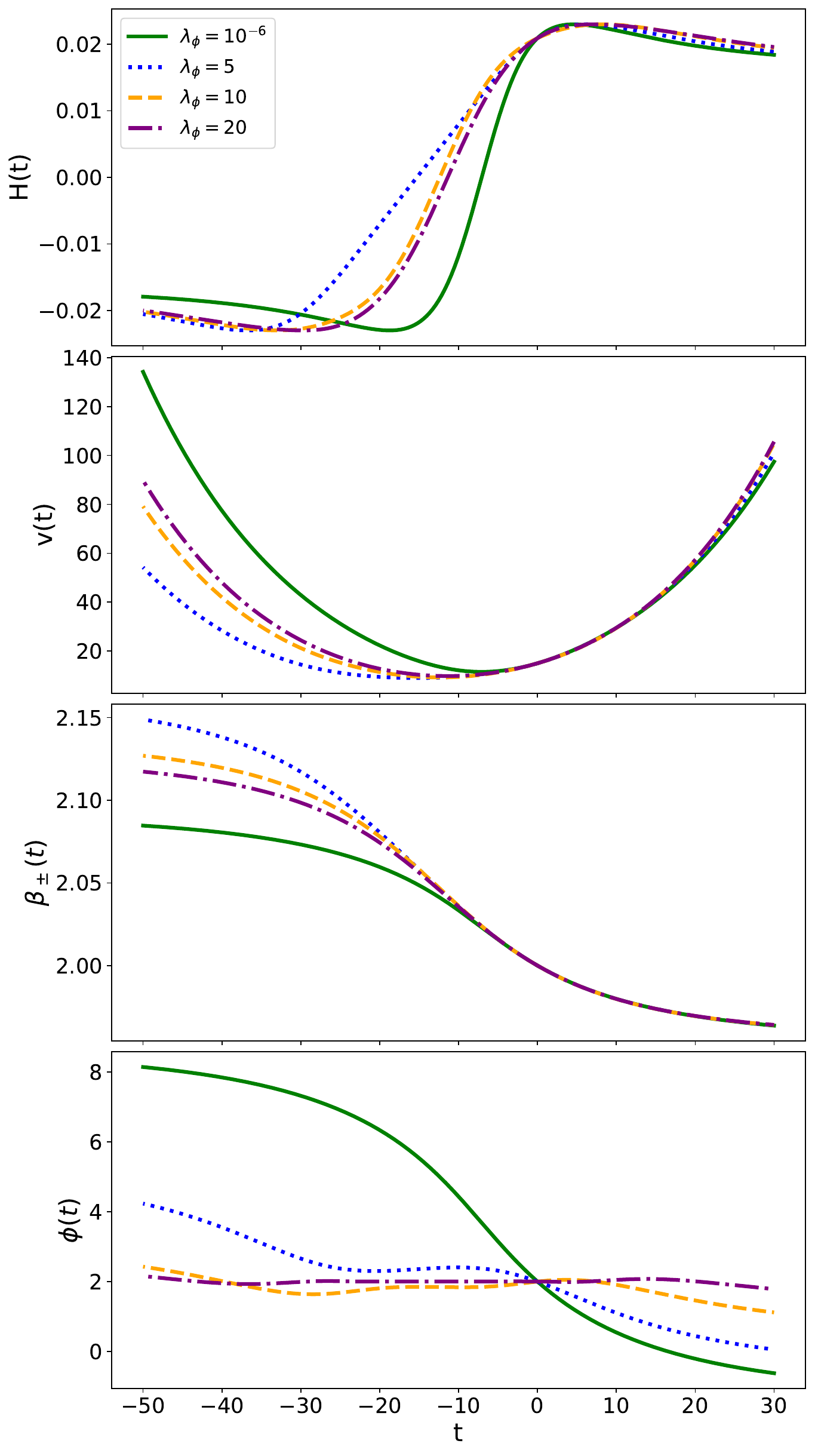}
 \caption{The effects of varying the matter polymer scale $\lp$ on the evolution of various cosmological quantities and the scalar field. The differences are clearly visible. The green (solid) curve shows the Schrodinger limit.}
 \label{fig-all-lambda-phi}
\end{figure}

\begin{figure}
\begin{center}
\includegraphics[scale=0.4]{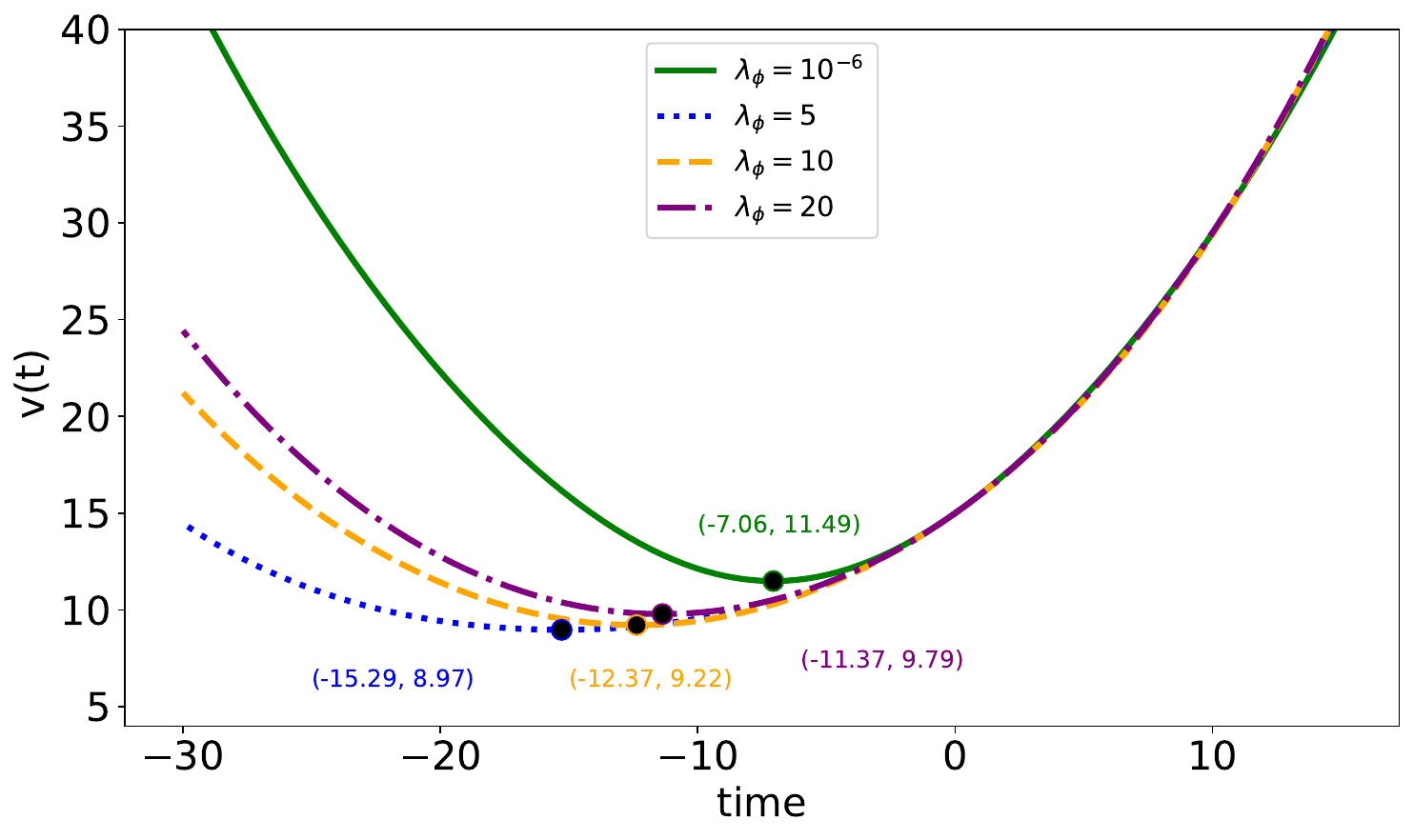}
\end{center}
\caption{\label{fig-lpzoom} Dependence of the bounce on the matter polymer scale $\lp$. The circles show the point of the bounce, and the numerical values indicate the time and the minimum volume at the bounce respectively $(t_{\text{bounce}}, v_{\text{bounce}}).$ The green (solid) curve shows the Schrodinger limit.}
\end{figure}

\begin{figure}
\begin{center}
\includegraphics[scale=0.4]{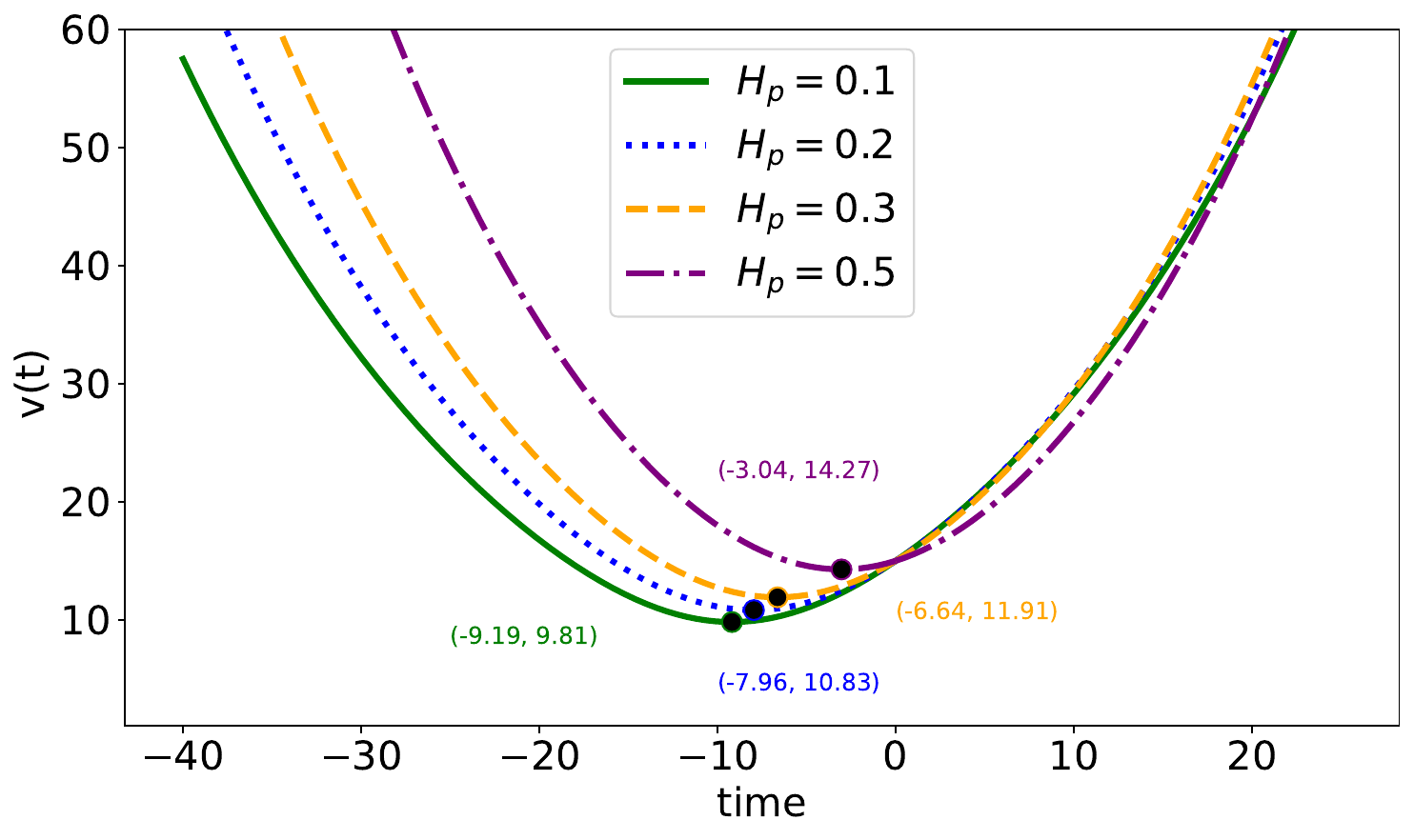}
\end{center}
\caption{\label{fig-IC}  Variation in the bounce as the value of the total Hamiltonian $H_p$ is varied. The circles show the point of the bounce, and the numerical values indicate the time and the minimum volume at the bounce respectively $(t_{\text{bounce}}, v_{\text{bounce}}).$ This shows dependence on initial conditions.}
\end{figure}

\begin{figure}
\begin{center}
\includegraphics[scale=0.35]{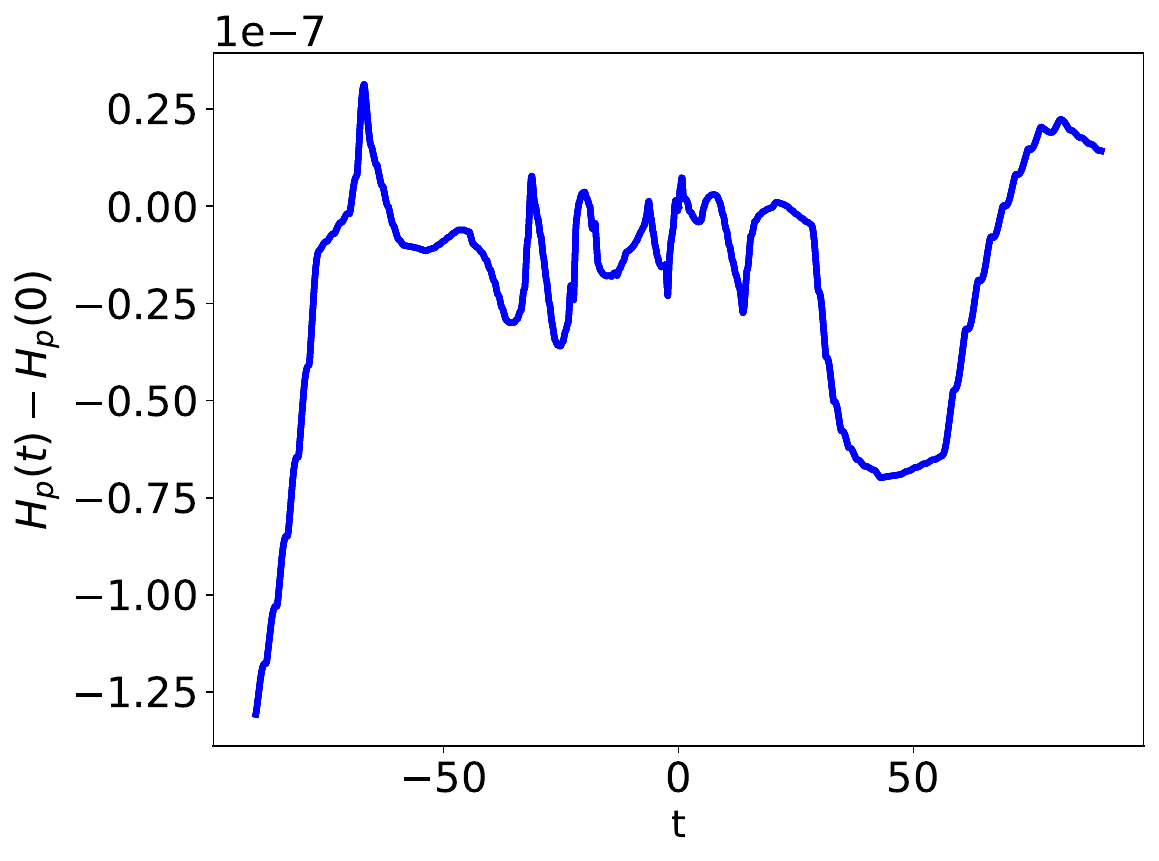}
\end{center}
\caption{\label{fig-AsymmetryHamil}Time variation of the physical Hamiltonian for a typical solution. The small scale of variations indicates good control over numerics.}
\end{figure}

We now consider the full system of differential equations (\ref{gra+sca}), which govern polymer Bianchi-I coupled to a polymer scalar field, and solve them numerically to analyze the resulting dynamics. Figure \ref{fig:all_three1} shows a typical evolution, and the effect of varying the anisotropy parameters. For all the remaining Figures (\ref{fig:all_three1} - \ref{fig-shearlp}), initial conditions were chosen to be $(v,p_v)|_{t=0}=(15, 1),$ $(\beta_\pm,p_\pm)|_{t=0}=(2,1),$ and $(\phi,p_\phi)|_{t=0} = (2,3)$ with parameter values $(\lv,\lambda_+, \lambda_-,\lambda_\phi)=(1,1,1,1),$ $\Lambda=0.07$ and $(\sigma_v,\sigma_+,\sigma_-,\sigma_\phi)=(1,1,1,1)$. If any parameter values differ, or were varied, they are indicated in the captions. Again, in the last panel of this figure, we note a similar oscillatory behavior in the Hubble parameter and anisotropy as seen in Figure \ref{fig:all_three}, occurring for large values of the anisotropy momentum.

Figure \ref{fig-all-lambda-v} shows the effects of varying the volume polymer scale $\lambda_v$, while keeping the matter and the anisotropy polymer scales fixed at $\lambda_\pm = \lp = 1$. The three distinct cases are $\lambda_v < \lambda_\pm$, $\lambda_v = \lambda_\pm$, and $\lambda_v > \lambda_\pm$, indicating situations where polymerization in the volume variable happens before, at the same scale, and after (respectively) polymerization in the anisotropies (and the scalar field). Given the construction of our model, the first two cases are physically reasonable.

Figure \ref{fig-all-lambda-phi} illustrates the effect on the dynamical behavior of the Hubble parameter, volume, anisotropies and the scalar field when the matter polymer scale $\lp$ is varied. We notice that there is a clear difference in the dynamics of the universe, in both the volume, and the anisotropies. In particular, the dynamics of the limiting case $\lp = 10^{-6}$ -- which represents a Schrodinger quantization of the scalar field -- are significantly different from when the field is polymer quantized. This shows that polymer quantization of the matter sector can have a significant effect on the gravitational observables.

We also note that there is variation in the location of the quantum bounce as the matter polymer scale, and the initial conditions are varied. These are illustrated in Figures \ref{fig-lpzoom} and \ref{fig-IC} respectively (to describe the effect of changing the initial conditions, we fix different values for the physical Hamiltonian $H_p,$ and solve for $p_v$). 

Lastly, Figure \ref{fig-AsymmetryHamil} shows the dynamics of the physical Hamiltonian after subtracting its initial value for a typical solution. Since, in dust time, the physical Hamiltonian $H_p$ is a constant of the motion, its value should remain unchanged under evolution. We find that this is indeed the case, and serves as a good check of our numerics.

\subsection{Directional scale factors, Hubble parameters, and shears}

\begin{figure}
    \centering
    \begin{subfigure}[t]{0.315\textwidth}
        \centering
        \includegraphics[width=\linewidth]{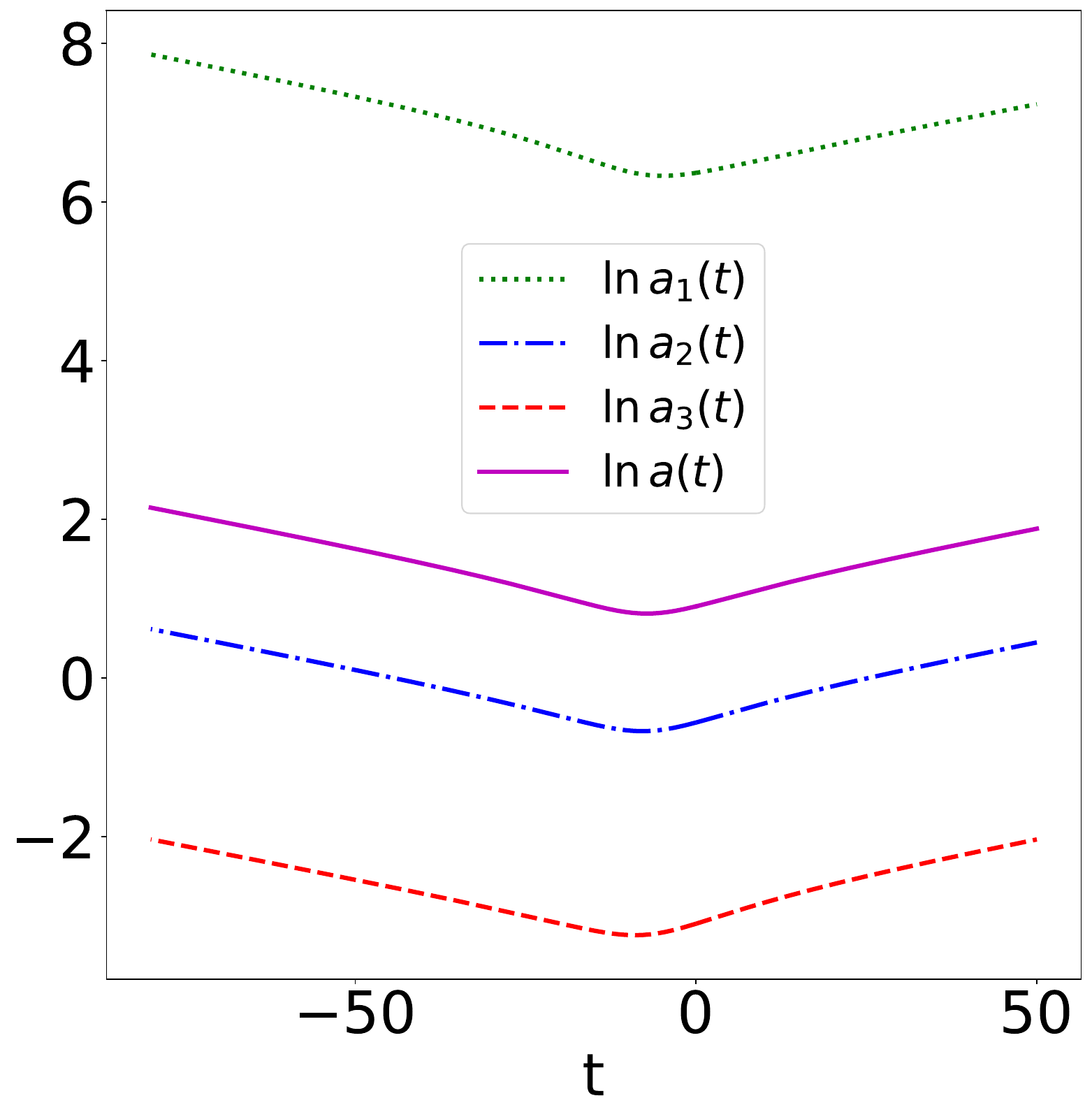}
        \label{fig-gvaryh17}
    \end{subfigure}%
     \hfill
    \begin{subfigure}[t]{0.335\textwidth}
        \centering
        \includegraphics[width=\linewidth]{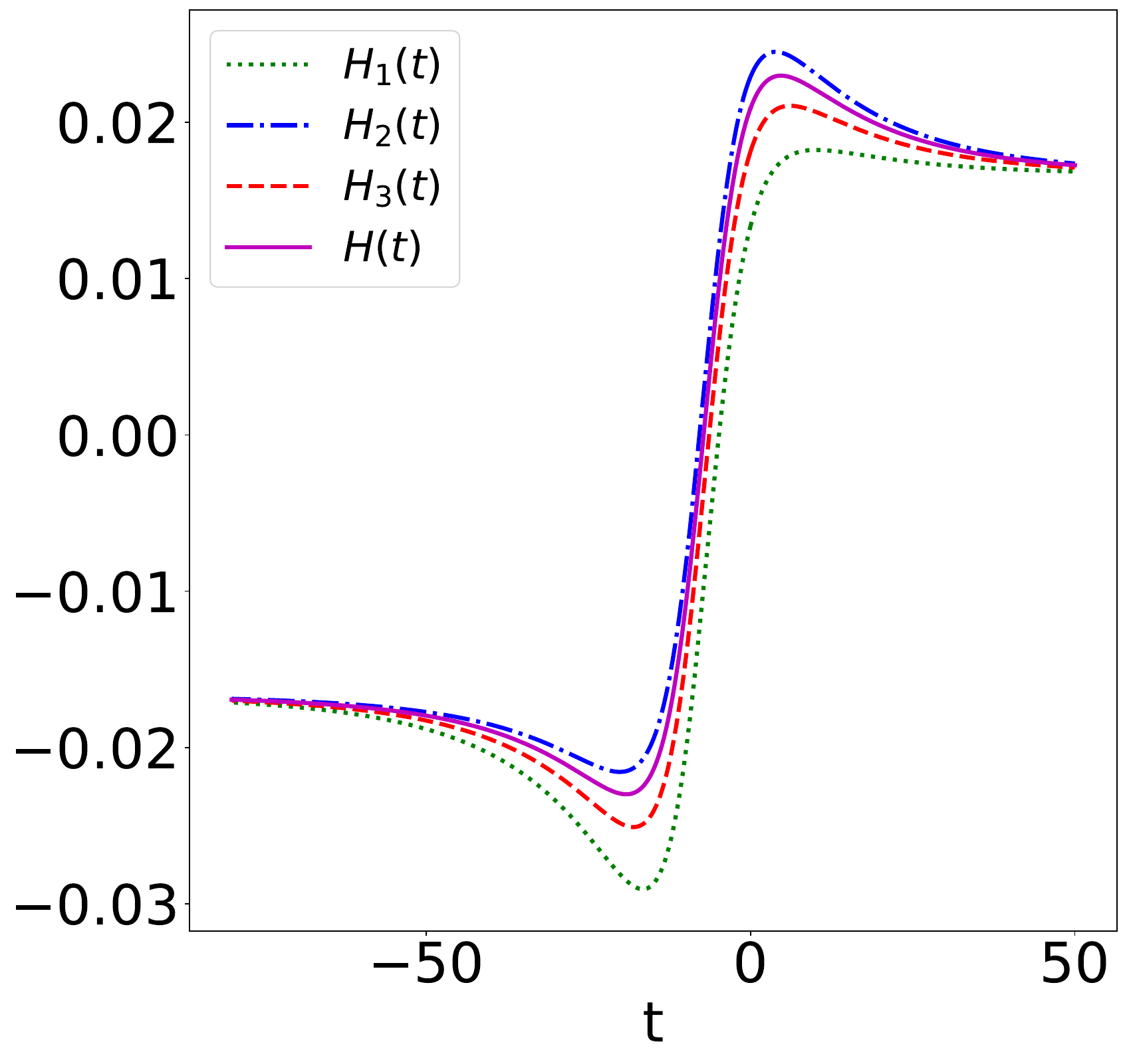}
        \label{fig-gvaryh19}
    \end{subfigure}%
     \hfill
    \begin{subfigure}[t]{0.32\textwidth}
        \centering
        \includegraphics[width=\linewidth]{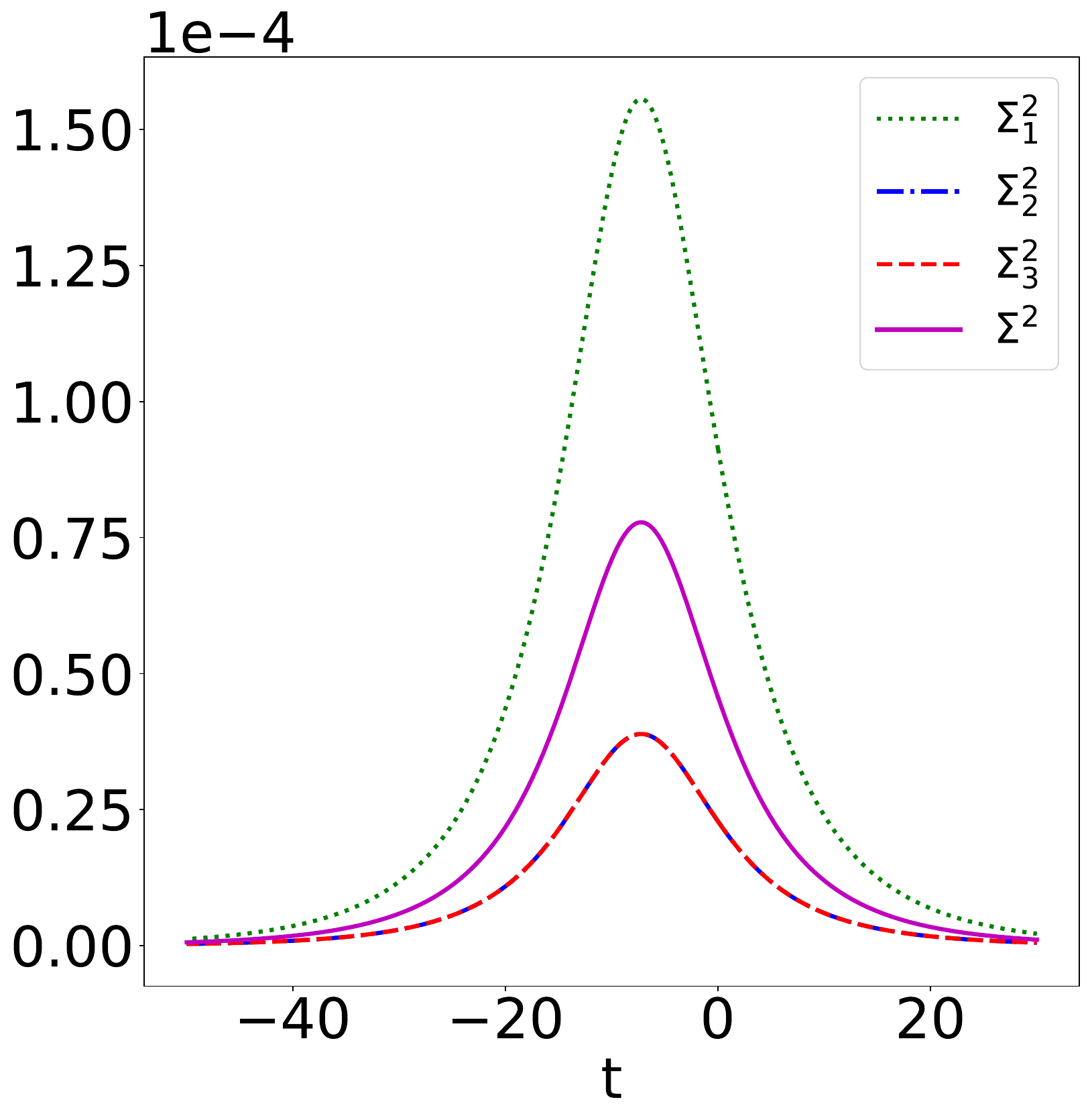}
       \label{fig-gvaryh211} 
    \end{subfigure}
    \caption{The three directional scale factors, Hubble parameters, shears, and their corresponding means.}
    \label{fig-all-lambda-factor}
\end{figure}

We have been looking at the dynamical trajectories of the Misner variables. However, once we have solved the equations of motion for the polymerized variables $v, \beta_\pm$ (\ref{gra+sca}), we can reconstruct the dynamical trajectories of the usual scale factors, the Hubble parameters, and the shears. We compute these trajectories from the classical relations given in (\ref{relation}), (\ref{direc hubble}, \ref{mean hubble}) and (\ref{direc shear}, \ref{shear1}) respectively (we note that these relations are not modified by polymerization, as they correspond to geometric definitions of observables. Polymer effects enter through the modified dynamics of the variables $v, \beta_\pm$). The scale factors, Hubble parameters, and shears, for a typical solution, are shown in Figure \ref{fig-all-lambda-factor}.

\begin{figure}
\begin{center}
\includegraphics[scale=0.3]{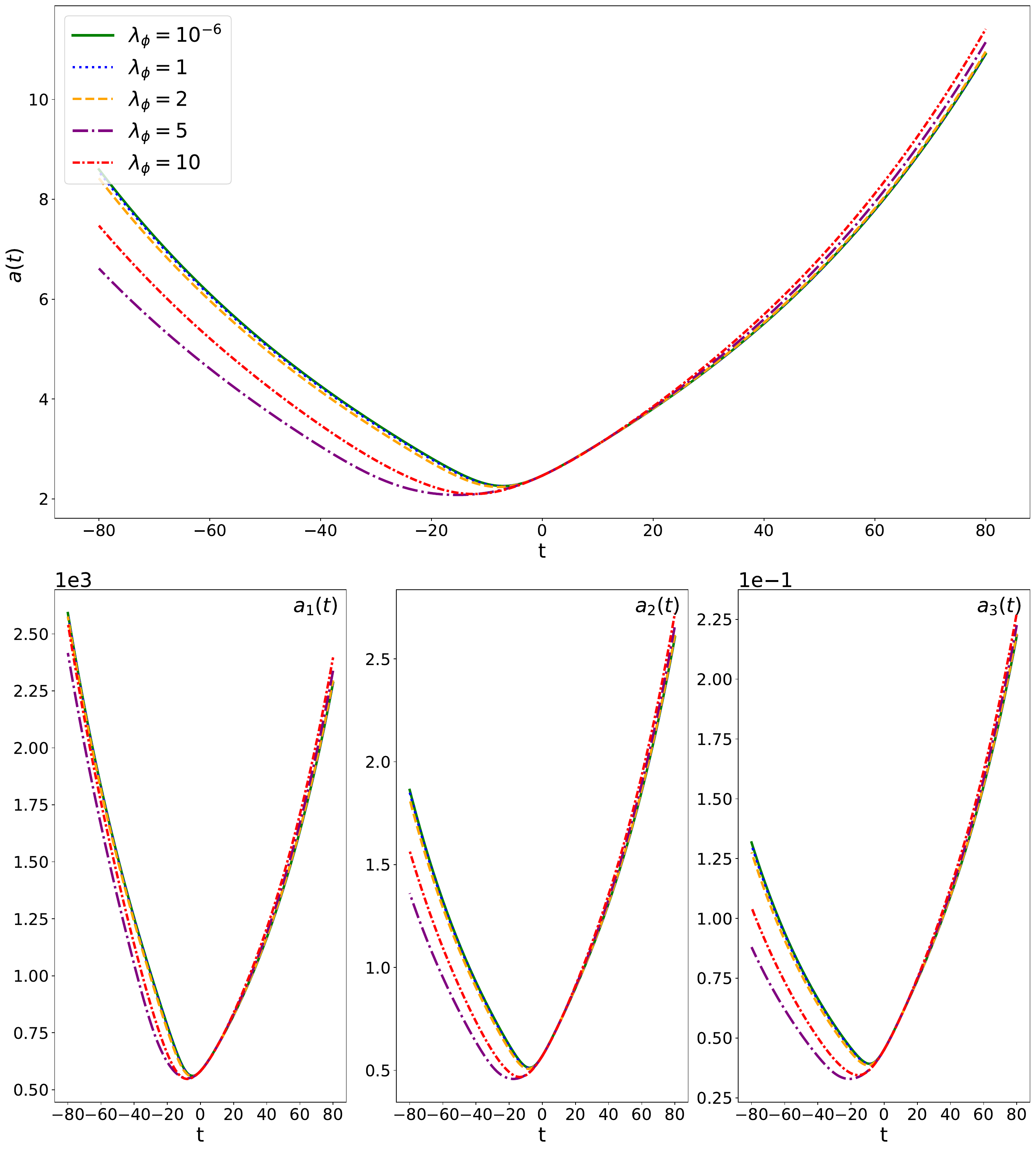}
\end{center}
\caption{\label{fig-scalelp} Variation in the three directional scale factors and their mean as $\lp$ is varied. The green (solid) curves show the Schrodinger limit.}
\end{figure}

\begin{figure}
\begin{center}
\includegraphics[scale=0.3]{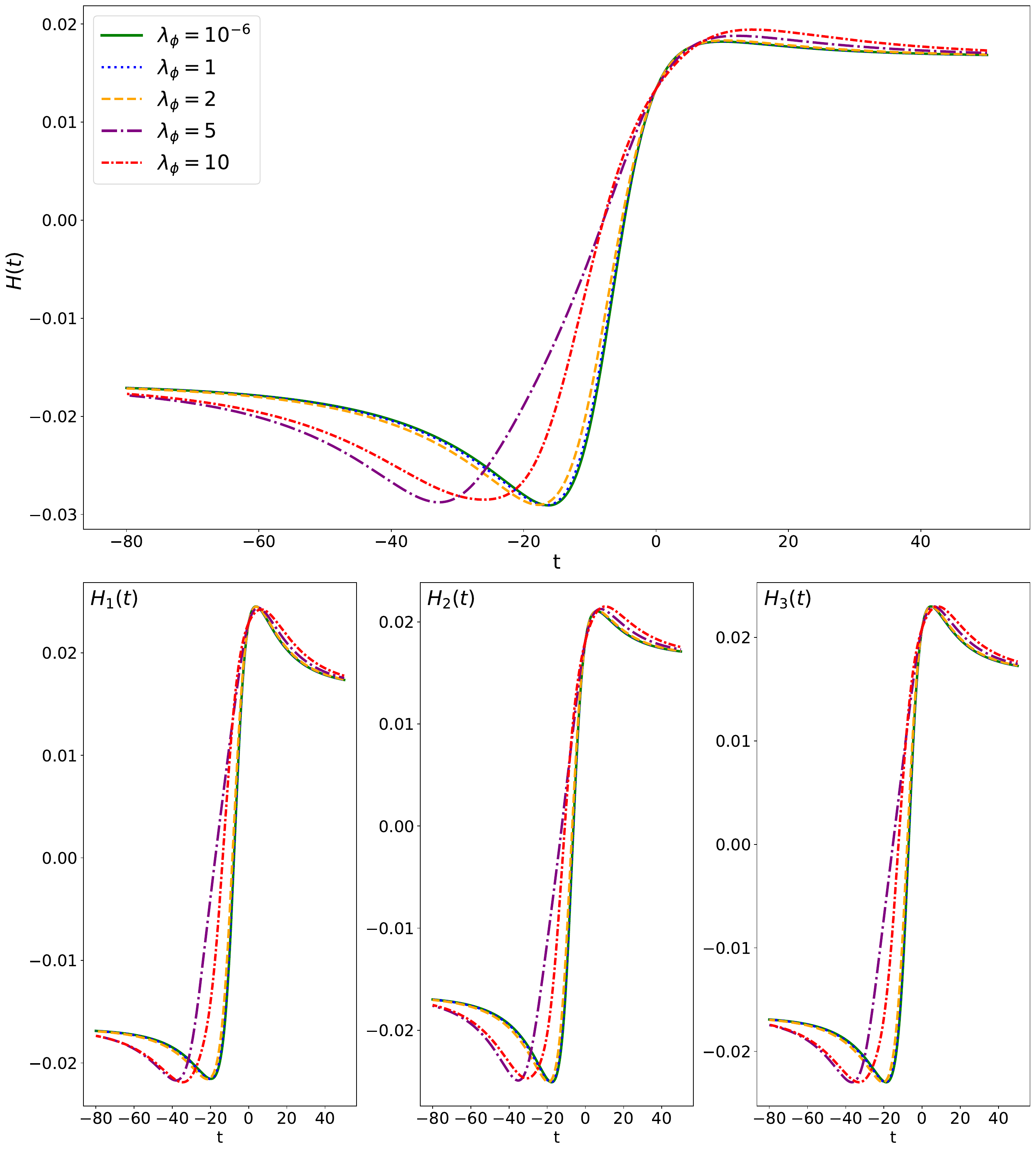}
\end{center}
\caption{\label{fig-direc hubble} Variation in the three directional Hubble parameters and their mean as $\lp$ is varied. The green (solid) curves show the Schrodinger limit. }
\end{figure}

\begin{figure}
\begin{center}
\includegraphics[scale=0.3]{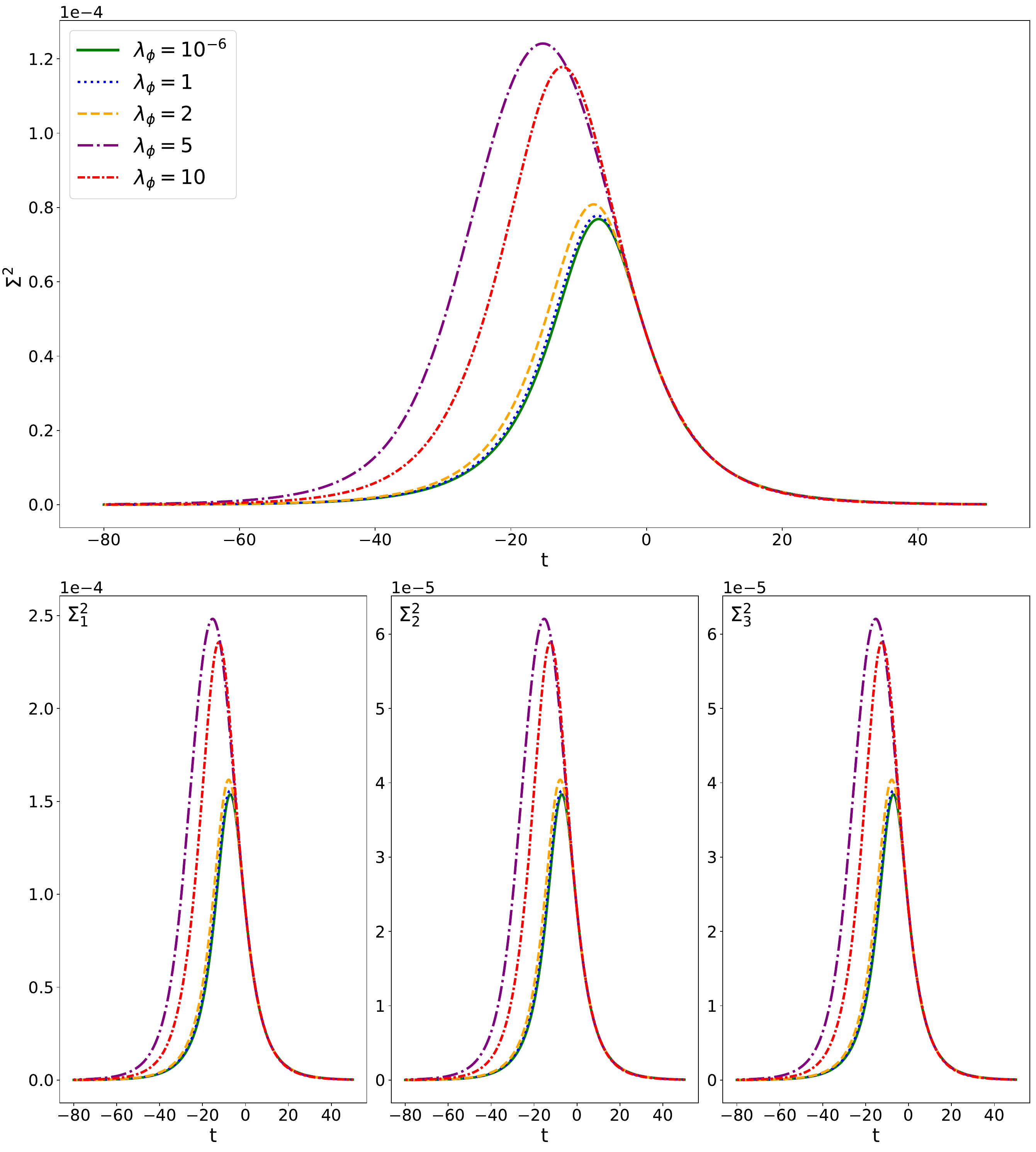}
\end{center}
\caption{\label{fig-shearlp} Variation in the three directional shears and their mean as $\lp$ is varied. The green (solid) curves show the Schrodinger limit.}
\end{figure}

The effects of varying the matter polymer scale $\lp$ on the dynamics of the scale factors (and their mean), the Hubble parameters (and their mean), and the shear scalars (and their mean) are shown in Figures \ref{fig-scalelp}, \ref{fig-direc hubble}, and \ref{fig-shearlp} respectively.

\subsection{Polymer scalar field effects}

To elucidate the effects arising solely from the polymer quantization of the scalar field (in addition to the dynamical differences discussed above), it is useful to simplify the equations of motion by taking all the states to be sharply peaked ($\sigma_v, \sigma_\pm, \sigma_\phi \rightarrow \infty$), fixing the parameter values $\lambda_v = \lambda_\pm = 1$ (these are the values we used in the numerics), and setting the anisotropy momenta (which are constants of motion) to one. With this, the $\phi$ and $p_v$ equations -- which are the only equations to contain the scalar field polymerization terms directly -- become,
\be
\dot{\phi} = \left( -\dfrac{\pp}{v} \right) F(2\Theta), ~~ \dot{p}_v = \left( -\dfrac{\pp^2}{2v^2} \right) G(\Theta) + g(v,p_v),
\ee
with,
\be
\Theta \equiv \dfrac{\lp \pp}{v}, ~~ F(\Theta) = \dfrac{\sin \Theta}{\Theta}, ~~ G(\Theta) = \dfrac{\sin 2 \Theta}{\Theta} - \left( \dfrac{\sin \Theta}{\Theta} \right)^2,
\ee
and $g(v,p_v)$ being the standard terms arising from the gravitational sector. In these simplified equations, the terms in the parentheses are the standard (Schrodinger) scalar field terms, and $F(\Theta), G(\Theta)$ represent terms arising from the polymerized scalar field. Note that in the Schrodinger limit ($\lp \rightarrow 0$), we have $F,G \rightarrow 1$ as expected.\footnote{$F,G$ also approach 1 at late times (as $v \rightarrow \infty$).} The lowest order polymer corrections appear at $\mathcal{O}(\lp^2)$, where $F \sim 1-2\Theta^2/3$ and $G \sim 1-\Theta^2$. Hence, polymerizing the scalar field produces deviations from the standard scalar field, which have observable effects on the dynamics.

\begin{figure}
    \centering
    \begin{subfigure}[t]{0.45\textwidth}
        \centering
        \includegraphics[width=\linewidth]{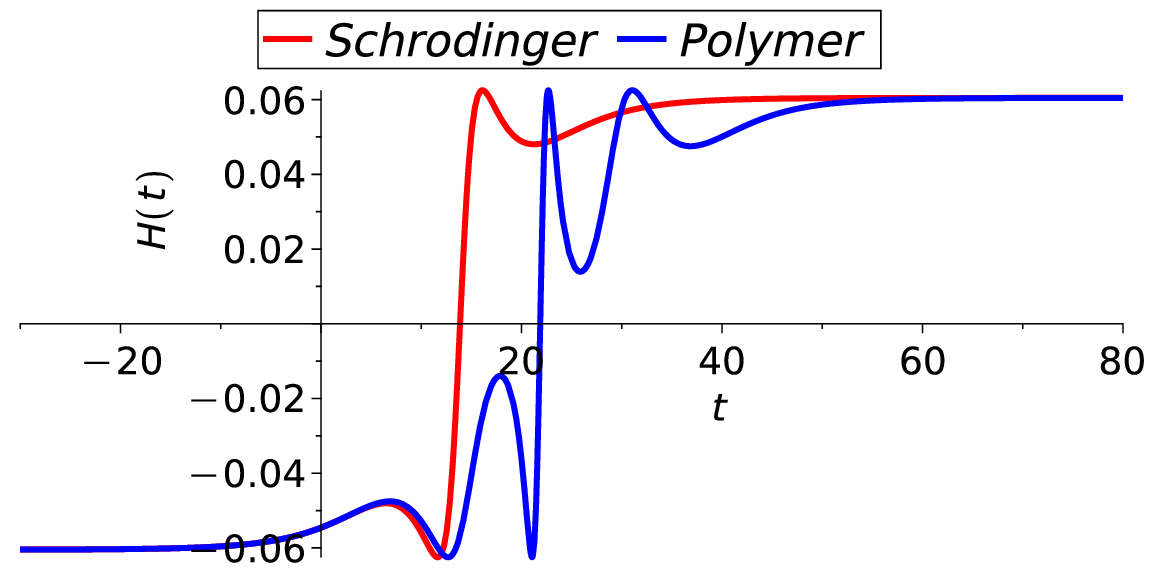}
    \end{subfigure}%
     \hfill
    \begin{subfigure}[t]{0.45\textwidth}
        \centering
        \includegraphics[width=\linewidth]{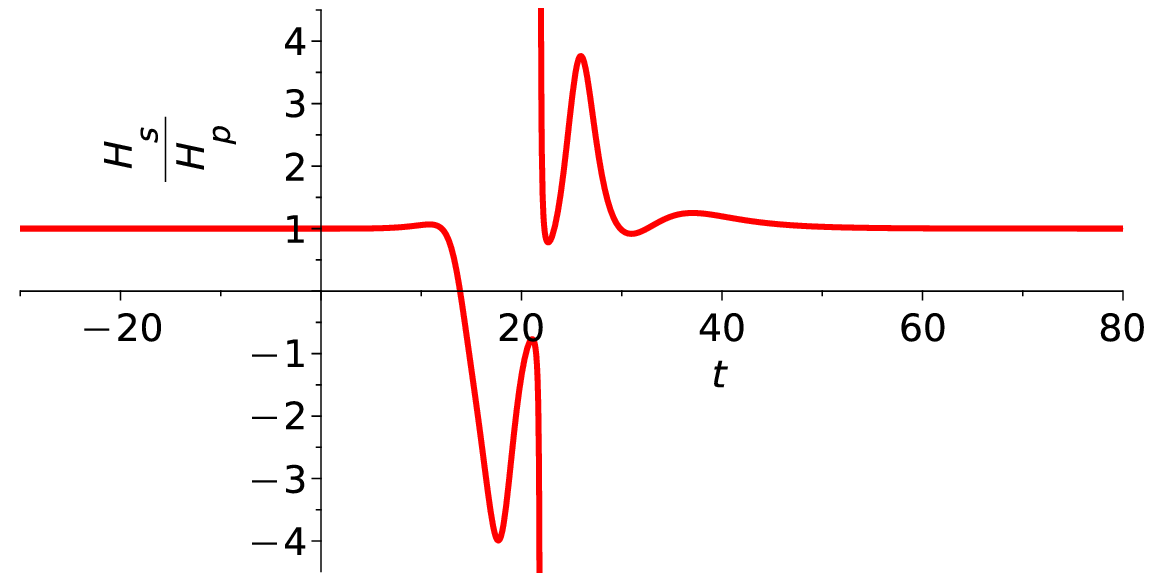}
    \end{subfigure}
    \caption{A comparison of the Hubble parameters for the Schrodinger and polymer cases. Left: The two Hubble factors. Right: Their ratio.}
    \label{fig-hubb-comp}
\end{figure}

Figure \ref{fig-hubb-comp} shows a comparison between the Hubble parameters for the Schrodinger and the polymer cases (with $\lp=0$ for the Schrodinger case, and $\lp = 1$ for the polymer case). For both cases, same parameter values and initial conditions were used (since bounce time depends on $\lp$, bounces for the two occur at different times). In the left panel, we see the two Hubble parameters, and the right panel shows their ratio (the vertical asymptote is the point where the polymer Hubble factor equals zero). There are clear differences in their evolutions near the bounce point, and at late times (as the universe expands) these differences become negligible and we recover the usual (Schrodinger) dynamics.

\begin{figure}
    \centering
    \begin{subfigure}[t]{0.45\textwidth}
        \centering
        \includegraphics[width=\linewidth]{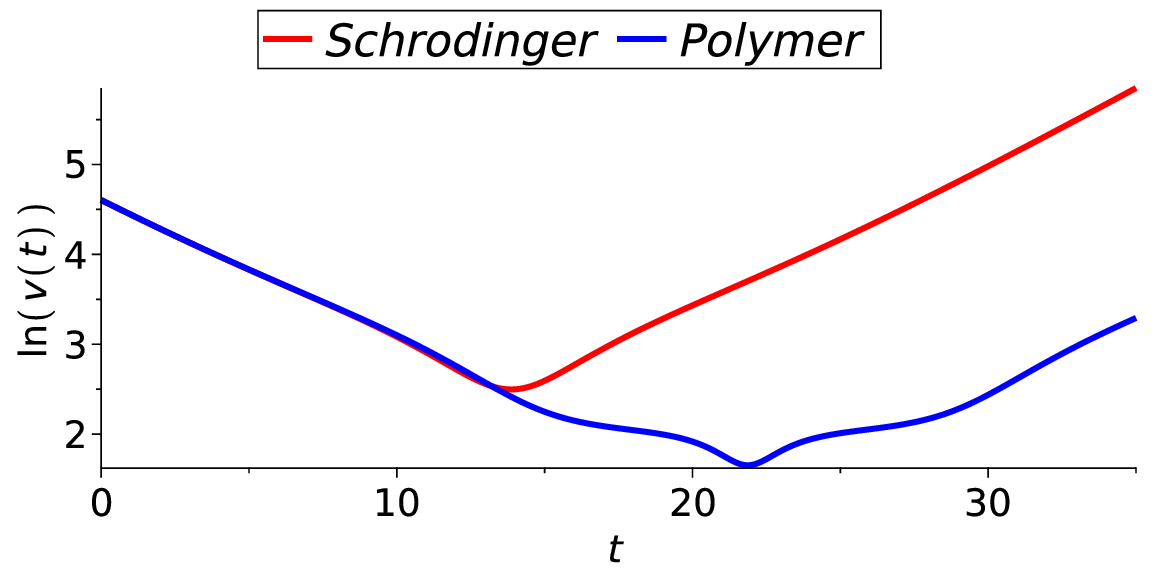}
    \end{subfigure}%
     \hfill
    \begin{subfigure}[t]{0.45\textwidth}
        \centering
        \includegraphics[width=\linewidth]{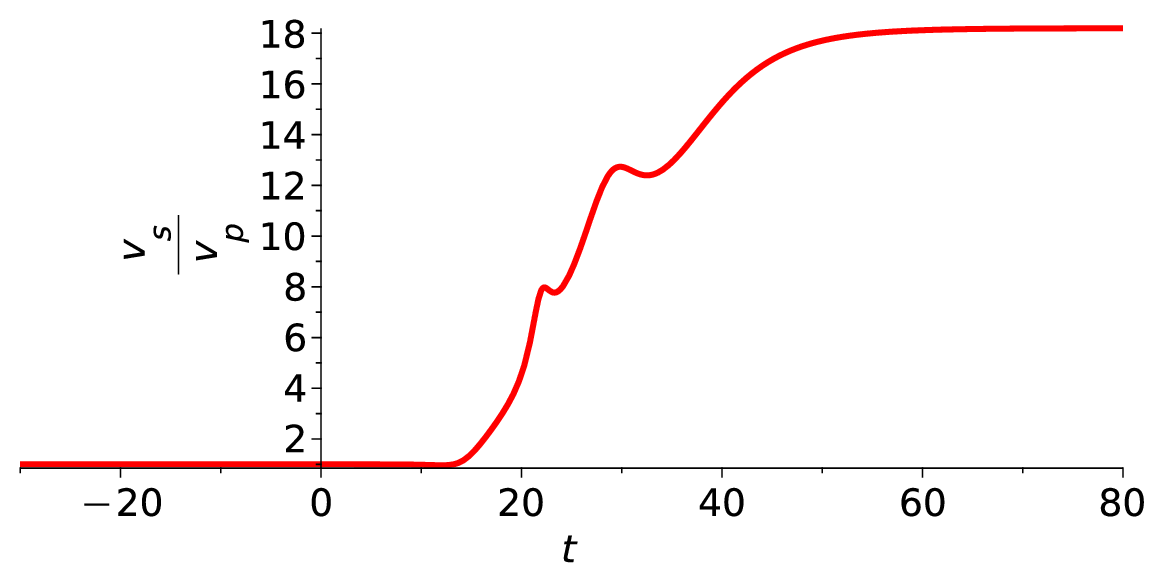}
    \end{subfigure}
    \caption{A comparison of the volumes for the Schrodinger and polymer cases. Left: (log of) The two volumes. Right: Their ratio.}
    \label{fig-vol-comp}
\end{figure}

\begin{figure}
    \centering
    \begin{subfigure}[t]{0.45\textwidth}
        \centering
        \includegraphics[width=\linewidth]{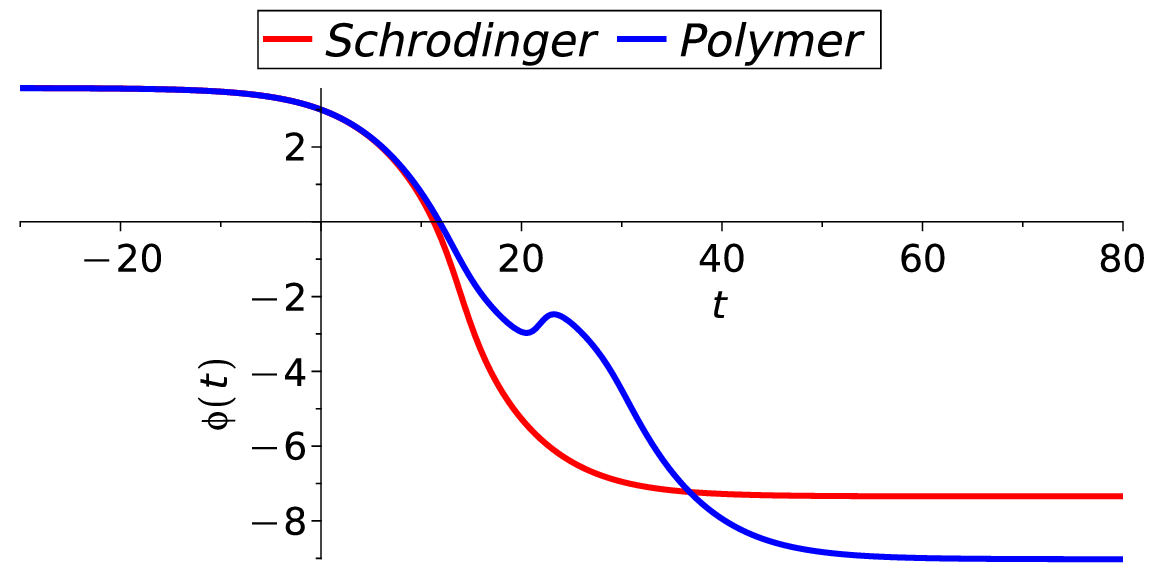}
    \end{subfigure}%
     \hfill
    \begin{subfigure}[t]{0.45\textwidth}
        \centering
        \includegraphics[width=\linewidth]{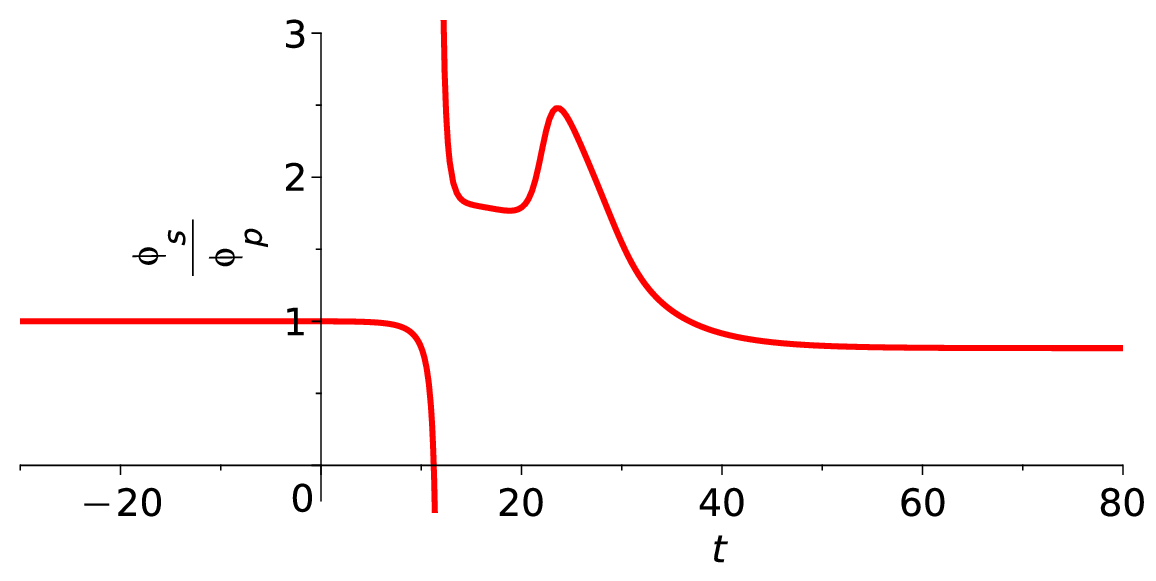}
    \end{subfigure}
    \caption{A comparison of the scalar fields for the Schrodinger and polymer cases. Left: The two scalar fields. Right: Their ratio.}
    \label{fig-phi-comp}
\end{figure}

Figures \ref{fig-vol-comp} and \ref{fig-phi-comp} show a similar comparison for the volume and the scalar field. Once again, we note that polymer effects only become significant near the bounce, where the polymerized version exhibits oscillations in both volume, and the scalar field. The differences in these dynamics can leave imprints on cosmological quantities which are in principle observable (for instance, a polymerized scalar field in an FLRW spacetime modifies the standard power spectrum \cite{Seahra:2012un}). We have shown here that such differences are indeed present (and are substantial near the bounce), however, a detailed analysis of the testable predictions arising from these differences warrants further study.

\section{Summary} \label{sec-con}

We have presented here a model of polymer quantization of a homogeneous but anisotropic Bianchi-I universe coupled to a polymer quantized massless scalar field. We use a pressureless dust field as an internal clock, and study the resultant effective dynamics obtained by taking the expectation value of the Hamiltonian operator in a semi-classical Gaussian state. This presents a complete (effective) quantization of both the gravity and the matter sectors, extending the results first obtained in the FLRW case \cite{Zulfiqar:2025aef} to an anisotropic universe.

We found that for a consistent polymer quantization of the anisotropies, we first have to polymerize the volume variable, compute an effective (volume) background, and then polymerize the anisotropies on that background (similar to what is done for the scalar field). This is because, in the Misner variables considered here, the anisotropy momenta are densities and not scalars. Whereas, to perform a polymer quantization, we have to consider variables constructed by exponentiating the momenta which requires them to be scalars. A related study of a Bianchi-I \emph{like} model where all gravitational momenta are scalars is currently under preparation \cite{Aleena-prep}.

Another possibility is to choose a different lapse function and rescale the Hamiltonian constraint such that all of the gravitational momenta are scalars (which, in our variables, corresponds to the choice $N=v^2$, and renders $p_{\alpha}$ and $p_{\pm}$ as scalars). This means, however, that dust can no longer be used as a clock, and a different choice of time gauge fixing needs to be performed (to remain consistent with a new choice of lapse). In general, such a choice will lead to a square root Hamiltonian which is challenging to quantize. Furthermore, in general, quantum theories with different clock choices are not equivalent (see, for example, \cite{Gielen:2021igw}). Alternatively, instead of fixing a (time) gauge classically, one can first quantize the theory, and then implement the constraint at the quantum level (as is more common in the literature). However, it is not clear if the two quantum theories are equivalent \cite{Schleich:1990gd}. Nevertheless, the primary advantage of such an approach would be that both the volume and the anisotropy momenta could be polymer quantized simultaneously. This would remove (time-dependent) factors of $v(t)$ from the exponential and cosine terms in the effective gravitational Hamiltonian (\ref{H_g}), giving rise to substantially different dynamics in general.

Our main results indicate that having different polymer scales, or state widths, for the two anisotropy variables gives rise to distinct cosmological dynamics. Similarly, depending on whether the volume polymer scale is larger than, equal to, or smaller than the anisotropy scale, the resultant dynamics are qualitatively different. Furthermore, the location of the bounce depends on initial conditions and the matter polymer scale. We also derived an effective Friedmann equation (which is not known yet for LQC models) that shows a quantum bounce, and we computed a relationship between the shear scalar and the effective anisotropy energy densities, similar in form to the effective Friedmann equation.

One of the main objectives of this study was to elucidate the effect of the matter polymer scale on the dynamics of the universe. We found that polymer quantization of the scalar field produces substantially different dynamics than a standard Schrodinger quantized, or classical, scalar field. Furthermore, we saw that varying the matter polymer scale leads to qualitatively different behavior for the volume and the anisotropies, as well as other cosmological quantities like the scale factors, the Hubble parameters, and the shear scalars, and it also affects the location of the quantum bounce.

A useful extension of this work would be to the Bianchi-IX (Mixmaster) model \cite{Misner:1969hg} where, as the singularity is approached, the geometry is approximated well by a Bianchi-I solution with (Kasner \cite{1921AmJM...43..217K}) transitions taking place between different Bianchi-I solutions \cite{Misner:1969ae}. A study of the semiclassical dynamics of this model (where only the gravity part is polymer quantized) appears in \cite{Muzammil:2025nuv}, and an analysis of the Mixmaster dynamics in the dust time gauge was discussed in \cite{Ali:2017qwa}. It will be interesting to see how the polymer quantization presented here (and in particular the polymerized matter field) will affect the Kasner transitions, and what its implications will be for the Belinski-Khalitnikov-Lifshitz (BKL) conjecture \cite{1971SvPhU..13..745B}. We leave this for future studies.

Overall, this project is a further step in the direction of analyzing the polymer quantized dynamics of gravity with matter where both of these sectors are treated on an equal footing (see also \cite{Ashtekar:2009mb, Ali:2022vhn, Han:2024ydv} for approaches where both sectors are quantized simultaneously). While we were limited here to discussing the effective dynamics that arise from this approach, and also with regards to the particular choice of polymer quantization we made, this study serves as a prelude towards a complete (polymer) quantum treatment of gravity with matter in general, and Bianchi-I cosmology with a scalar field in particular.

\backmatter

\bmhead{Acknowledgements}

SMH would like to thank Dr. Rizwan Khalid for helpful discussions and suggestions. This project was supported by the Higher Education Commission (HEC) of Pakistan through NRPU grant No. 20-15435.

\section*{Declarations}

\textbf{Data Availability Statement:} We do not analyse or generate any datasets, because our work proceeds within a theoretical and mathematical approach.

\bibliography{biblography}
 
\end{document}